\documentclass[12pt,english,eqsecnum,english,prd,aps,nofootinbib,superscriptaddress,amsfonts,longbibliography,tightenlines,showpacs]{revtex4-1}
\usepackage[T1]{fontenc}
\usepackage[utf8]{inputenc}
\usepackage{color}
\usepackage{babel}
\usepackage{mathtools}
\usepackage{amsmath}
\usepackage{amssymb}
\usepackage{stmaryrd}
\usepackage[unicode=true,pdfusetitle,
 bookmarks=true,bookmarksnumbered=false,bookmarksopen=false,
 breaklinks=true,pdfborder={0 0 1},backref=false,colorlinks=false]
 {hyperref}



\begin{document}
\title{Phenomenology of a massive quantum field in a cosmological quantum
spacetime}

\author{Saeed Rastgoo}
\email{srastgoo@yorku.ca}
\affiliation{Department of Physics and Astronomy, York University~\\
 4700 Keele Street, Toronto, Ontario M3J 1P3, Canada}
\affiliation{School of Sciences and Engineering, Monterrey Institute of Technology
(ITESM), Campus Le\'on~\\
 Av. Eugenio Garza Sada, Le\'on, Guanajuato 37190, Mexico}

\author{Yaser Tavakoli}
\email{yaser.tavakoli@guilan.ac.ir}
\affiliation{Department of Physics, University of Guilan, 41335-1914 Rasht, Iran}
\affiliation{Faculty of Physics, University of Warsaw, Pasteura 5, 02-093 Warsaw,
Poland}
\affiliation{Departamento de F\'isica, Universidade Federal do Esp\'irito Santo,~\\
 Av. Fernando Ferrari 514, Vit\'oria - ES, Brazil}

\author{Julio C. Fabris}
\email{fabris@pq.cnpq.br}
\affiliation{Departamento de F\'isica, Universidade Federal do Esp\'irito Santo,~\\
 Av. Fernando Ferrari 514, Vit\'oria - ES, Brazil}
\affiliation{National Research Nuclear University ``MEPhI'', Kashirskoe sh. 31,
Moscow 115409, Russia}
\date{\today}

\begin{abstract}
We revisit the quantum theory of a massive, minimally coupled scalar
field, propagating on the Planck-era isotropic cosmological quantum
spacetime which transitions to a classical spacetime in later times.
The quantum effects modify the isotropic spacetime such that effectively
it exhibits anisotropies. Thus, the interplay between this quantum
background and modes of the field, when disregarding the backreactions,
gives rise to a theory of a quantum field on an anisotropic, dressed
spacetime. Different solutions are found
whose components depend on the  quantum fluctuations
of the background geometry. We  construct a formal expression for the power
spectrum of the scalar field fluctuations on such anisotropic background.
It is shown that the anisotropy of this power spectrum is due to the
modified frequency of the propagating quantum modes. This provides
a quantitative estimate for the deviations from the isotropic power
spectrum which can lead to the potential observational signatures
on the cosmic microwave background. In addition, the problem of particle
production when transitioning from such an effective spacetime to
a classical one is reexamined. It is shown that particles are created,
and the expectation value of their number operator depends on the
quantum geometry fluctuations.
\end{abstract}
\maketitle

\section{Introduction}

The study of inhomogeneous, anisotropic Universe is of particular
interest in general relativity, in order to avoid postulating special
initial conditions as well as the existence of particle horizon in
isotropic models \cite{Misner:1967uu}. Based on the observational
evidence, the consensus of the scientific community is that the spatial
Universe and its expansion in time are quite isotropic. Hence, to
explain this transition from anisotropic very early Universe to the
late-time isotropic one, we may require a mechanism for damping down
the inhomogeneity and anisotropy. Furthermore, as is common in time
dependent geometries, it was suggested that this transition will give
rise to creation of particles \cite{Hu:1978zd,Zeldovich:1970si}.

One way to explain the anisotropies in the very early universe is
to associate their emergence to the quantum effects present in the
background spacetime in those early times. A suitable setting to do
such an analysis is loop quantum cosmology (LQC) \cite{Ashtekar:2008zu,Ashtekar:2003hd}
which is a cosmological theory inspired by Loop quantum gravity (LQG),
which itself is a background independent, non-perturbative approach
to quantization of general relativity\cite{Ashtekar:2004eh,Rovelli:2004tv,Thiemann:2007zz}.
In this framework, the quantum nature of the big bang has been investigated
for the isotropic FLRW models \cite{Ashtekar:2006rx,Ashtekar:2006uz,Ashtekar:2006wn,Ashtekar:2007em}
and the simplest anisotropic model \cite{Ashtekar:2009vc}. These
investigations show that the big bang singularity is resolved within
LQC, and is replaced by a quantum bounce at which the energy density
of the universe has a maximum critical value $\rho_{{\rm crit}}=0.41\rho_{{\rm Pl}}$,
of the order of Planck density $\rho_{{\rm Pl}}$ (see Ref. \cite{Ashtekar:2011ni}
for a review of the recent developments in LQC).

In order to explore the properties of the propagation of the quantum
matter fields on the effective background in the early universe, and
their behavior when the spacetime transitions to a classical one,
one needs to study these matter fields coupled to a spacetime geometry
which is quantized due to LQC. Such a quantum theory of test fields
propagating on cosmological quantum spacetimes has been investigated
in the presence of an isotropic, flat FLRW \cite{Ashtekar:2009mb}
and the anisotropic Bianchi type-I \cite{Dapor:2012jg} background
geometries. There, by using a scalar field $\phi$ as a relational
time parameter (or clock variable), the Hamiltonian constraint is
deparametrized, which allows for a description of the evolution of
the universe in terms of this physical time parameter. The field $\phi$
used as a clock variable is called the background matter source (as
opposed to other scalar ``test fields'' $\varphi$ present in the
theory). This type of analysis was generalized in Ref. \cite{Dapor:2012qn},
to a case where an irrotational dust was introduced for the physical
time variable in quantum theory. The main result of both investigations
is that, an effective (semiclassical) background spacetime would emerge
on which the fields propagate, and whose metric components depend
on the fluctuations of the quantum geometry operators. Some interesting
phenomenological features of these effective geometries were studied
in \cite{Dapor:2012jg,Dapor:2012qn,Assaniousssi:2014ota}. A significant
extension of Ref. \cite{Ashtekar:2009mb} was recently performed,
in order to generalize the standard theory of cosmological perturbations
to include the Planck regime \cite{Agullo:2012fc}. The strategy
therein was to truncate the classical general relativity coupled to
a scalar field, to a sector including homogeneous, isotropic configuration,
together with the first order inhomogeneous perturbations, and the
quantum theory of the truncated phase space was constructed using
the techniques of LQG. This framework was applied to explore pre-inflationary
dynamics of the early universe \cite{Agullo:2013ai,Ashtekar:2015dja}.
Another framework was provided in Ref. \cite{Dapor:2013pka} for
quantization of linear perturbations on a quantum background spacetime,
by introduction of a different choice of fundamental variables to
those that are usually used in quantizing the perturbations on a fixed
classical background. In particular, in the presence of a natural
gauge-fixing in the theory, the old variables could be used as fundamental
operators, which provided a true dynamics in terms of the homogeneous
part of the relational time.

In the present paper, using LQC, we consider the effects corresponding
to the propagation of a test field on a quantized early universe background,
and the associated particle production when this effective spacetime
transitions to a classical one. In section \ref{QFT-LQG}, we show
that, in the presence of a test field, the effective spacetime resembles
an anisotropic spacetime due to quantum gravity effects. Then, we
consider several cases where either the mass of the test field or
the geometry (i.e. its associated scale factors) are dressed and some
of the consequences. In section \ref{pairprod}, we examine the phenomena
of particle creation in transitioning from the effective anisotropic
spatially homogeneous background to a classical isotropic one. In
section \ref{fluctuation}, we discuss the effects that emergent anisotropies
can have on the power spectrum of the CMB and its associated possible
observable signatures. Finally, in section \ref{conclusion} we will
conclude and discuss some of the outlooks of our work.

\section{Quantum fields in quantum spacetimes\label{QFT-LQG}}

In this section, we study the quantum theory of a test field $\varphi$,
both massless and massive, on a \emph{quantized FLRW background},
and compare it to the behavior of the same test field on a \emph{classical
anisotropic Bianchi I metric}. In both cases, the background geometry
is coupled to a massless scalar field $\phi$, which plays the role
of the physical internal time. Hence, the background elements consist
of the background spacetime and the massless field $\phi$ as the
clock variable, while the test field $\varphi$ is propagating on
this background spacetime. 

We will see that the resulting evolution (Schrödinger) equations for
both cases bare a striking resemblance to each other, an observation
that leads us to the rest of our analysis about the several possible
scenarios about the behavior of the matter field $\varphi$ on the
effective FLRW spacetime. To this end, we start with the classical
Bianchi I metric, and using that, work our way towards the quantum
FLRW model, since it is easier to go from a more general anisotropic
model to an isotropic one.

\subsection{Quantum matter propagating on a classical Bianchi I spacetime}

A Bianchi type-I spacetime, is represented by the anisotropic background
metric
\begin{equation}
g_{ab}dx^{a}dx^{b}=-N_{x^{0}}^{2}\left(x^{0}\right)\left(dx^{0}\right)^{2}+\sum_{i}^{3}a_{i}^{2}\left(x^{0}\right)\left(dx^{i}\right)^{2},\label{BI-metric-class}
\end{equation}
with $N_{x^{0}}$ being the lapse function and $a_{i}$ the scale
factors. The metric \eqref{BI-metric-class} is written in coordinates
$(x_{0},\mathbf{x})$, in which $\mathbf{x}\in\mathbb{T}^{3}$ (3-torus
with coordinates $x^{j}\in(0,\ell_{j})$), and $x^{0}\in\mathbb{R}$
is a generic time coordinate. Later, we can set $x^{0}=\phi$ so that
the evolution becomes relational. Furthermore, we consider a real
(inhomogeneous), minimally coupled, free scalar test field $\varphi(x_{0},\mathbf{x})$,
with mass $m$, propagating on this background spacetime. The Hamiltonian
of the scalar test field can be written as the sum of the Hamiltonians
$H_{\mathbf{k}}\left(x^{0}\right)$ of the decoupled harmonic oscillators,
each of which written in terms of a pair $(\mathsf{q}_{\mathbf{k}},\mathsf{p}_{\mathbf{k}})$,
as \cite{Dapor:2012jg} 
\begin{equation}
H_{\varphi}(x^{0})\coloneqq\sum_{\mathbf{k}\in{\cal L}}H_{\mathbf{k}}(x^{0})=\frac{N_{x^{0}}(x^{0})}{2|a_{1}a_{2}a_{3}|}\sum_{\mathbf{k}\in{\cal L}}\Big(\mathsf{p}_{\mathbf{k}}^{2}+\omega_{k}^{2}(x^{0})\mathsf{q}_{\mathbf{k}}^{2}\Big).\label{Hamiltonian-SF1}
\end{equation}
Here, ${\cal L}={\cal L}_{+}\cup{\cal L}_{-}$ is a 3-dimensional
lattice spanned by $\mathbf{k}=(k_{1},k_{2},k_{3})\in(2\pi\mathbb{Z}/\ell)^{3}$,
with $\mathbb{Z}$ being the set of integers and $\ell^{3}\equiv\ell_{1}\ell_{2}\ell_{3}$
\cite{Ashtekar:2009mb,Dapor:2012jg}: 
\begin{equation}
\forall\mathbf{k}\in{\cal L}_{+}:{\cal L}_{+}\coloneqq\{k_{3}>0\}\cup\{k_{3}=0,k_{2}>0\}\cup\{k_{2},k_{3}=0,k_{1}>0\},\quad{\rm and}\ -\mathbf{k}\in{\cal L}_{-}.\label{lattice}
\end{equation}
The conjugate variables $\mathsf{q}_{\mathbf{k}}$ and $\mathsf{p}_{\mathbf{k}}$,
associated with the $\mathbf{k}$'th mode of the field satisfy the
relation $\{\mathsf{q}_{\mathbf{k}},\mathsf{p}_{\mathbf{k}^{\prime}}\}=\delta_{\mathbf{k},\mathbf{k}^{\prime}}$.
Moreover, $\omega_{k}\left(x^{0}\right)$ is a time-dependent frequency
which is defined by 
\begin{equation}
\omega_{k}^{2}\left(x^{0}\right)\coloneqq|a_{1}a_{2}a_{3}|^{2}\left[\sum_{i=1}^{3}\left(\frac{k_{i}}{a_{i}}\right)^{2}+m^{2}\right].\label{frequency-SF-b1}
\end{equation}
In other words, $\mathsf{q}_{\mathbf{k}}$ is the field amplitude
for the mode characterized by $\mathbf{k}$, satisfying the Klein--Gordon
equation $(\square-m^{2})\varphi=0$, which is the equation of motion
obtained from this Hamiltonian \eqref{Hamiltonian-SF1}.

Since the background is left as a classical one, one needs to only
quantize the test field. The quantization of a mode $\mathbf{k}$
of the test field $\varphi$ resembles that of a quantum harmonic
oscillator with the Hilbert space {\small{}${\cal H}_{\varphi}^{(\mathbf{k})}=L_{2}(\mathbb{R},d\mathsf{q}_{\mathbf{k}})$}.
The canonical variables are promoted to operators on this Hilbert
space as $\hat{\mathsf{q}}_{\mathbf{k}}\psi(\mathsf{q}_{\mathbf{k}})=\mathsf{q}_{\mathbf{k}}\psi(\mathsf{q}_{\mathbf{k}})$
and $\hat{\mathsf{p}}_{\mathbf{k}}\psi(\mathsf{q}_{\mathbf{k}})=-i\hbar\partial/\partial\mathsf{q}_{\mathbf{k}}\psi(\mathsf{q}_{\mathbf{k}})$,
and the time evolution of any state $\psi(\mathsf{q}_{\mathbf{k}})$
is generated by the Hamiltonian operator $\hat{H}_{\mathbf{k}}$ via
the Schrödinger equation 
\begin{equation}
i\hbar\partial_{x^{0}}\psi(x^{0},\mathsf{q}_{\mathbf{k}})=\frac{\ell^{3}N_{x^{0}}}{2V}\Big(\hat{\mathsf{p}}_{\mathbf{k}}^{2}+\omega_{k}^{2}\hat{\mathsf{q}}_{\mathbf{k}}^{2}\Big)\psi(x^{0},\mathsf{q}_{\mathbf{k}}),\label{Hamiltonian-SF-bquantum}
\end{equation}
where $V$ is the physical volume\footnote{As in LQC of Bianchi I model, we fix a fiducial cell ${\cal V}$,
and take its edges to lie along the integral curves of the fiducial
triad, with coordinate lengths $\ell_{1},\ell_{2},\ell_{3}$, so that
the volume of ${\cal V}$ is $\mathring{V}=\ell_{1}\ell_{2}\ell_{3}\equiv\ell^{3}$.
Then, after parametrization of the gravitational phase space by a
pair $(c^{i},p_{i})$, the physical volume reads $V=\sqrt{p_{1}p_{2}p_{3}}=\ell^{3}|a_{1}a_{2}a_{3}|$.
Note that, the relation between the phase space variables is given
by $p_{i}\equiv\epsilon_{ijk}\ell_{j}\ell_{k}a_{j}a_{k}$ ($\epsilon_{ijk}$
is the Levi-Civita symbol) \cite{Ashtekar:2009vc}.} of the universe which is given by 
\begin{equation}
V=\ell^{3}\left|a_{1}a_{2}a_{3}\right|.\label{eq:V-univ}
\end{equation}
Denoting the Bianchi I variables with a tilde, using \eqref{frequency-SF-b1}
and \eqref{eq:V-univ} in \eqref{Hamiltonian-SF-bquantum}, and setting
$x^{0}=\phi$, one can see that the evolution of the quantum state
$\psi$ with respect to the internal physical time $\phi$ for a mode
$\tilde{\mathbf{k}}$ of a test field on a \emph{classical Bianchi
I background} \eqref{BI-metric-class} is described by the Schrödinger
equation
\begin{equation}
i\hbar\partial_{\phi}\psi\left(\phi,\mathsf{q}_{\mathbf{k}}\right)=\frac{\tilde{N}_{\phi}}{2\left|\tilde{a}_{1}\tilde{a}_{2}\tilde{a}_{3}\right|}\left[\hat{\mathsf{p}}_{\mathbf{k}}^{2}+\left(\sum_{i}^{3}\frac{\tilde{k}_{i}^2}{\tilde{a}_{i}^2}+\tilde{m}^{2}\right)\left(\tilde{a}_{1}\tilde{a}_{2}\tilde{a}_{3}\right)^{2}\hat{\mathsf{q}}_{\mathbf{k}}^{2}\right]\psi\left(\phi,\mathsf{q}_{\mathbf{k}}\right).\label{Hamiltonian-SF-bquantum-eff}
\end{equation}
where $\tilde{a}_{i}$'s (with $i=1,2,3$) are now functions of the
relational time $\phi$, i.e., $\tilde{a}_{i}=\tilde{a}_{i}(\phi)$.
This is the evolution equation of a quantum test field propagating
on a classical Bianchi I geometry, which we will be comparing to the
one with the test field propagating on a quantum FLRW spacetime, which
will be derived in the next subsection. 

\subsection{Quantum matter propagating on a quantum FLRW spacetime\label{backreaction}}

Now we will consider the propagation of a \emph{quantum test field
over a quantum FLRW spacetime}. Using the analysis of the previous
subsection, one can go to an isotropic regime by setting $a_{1}=a_{2}=a_{3}\equiv a\left(x^{0}\right)$.
If, for this case, we choose a harmonic time coordinate, $x^{0}=\tau$,
then the corresponding lapse $N_{\tau}$ will be related to the lapse
$N_{\phi}$ via
\begin{align}
N_{\phi}= & \left(\frac{\ell^{3}}{\tilde{p}_{\phi}}\right)N_{\tau},\label{eq:N-phi-tilde}\\
N_{\tau}= & \left|a_{1}\left(\phi\right)a_{2}\left(\phi\right)a_{3}\left(\phi\right)\right|.\label{eq:N-tau-tilde}
\end{align}
Given the isotropic nature of the background in this case, the lapse
function becomes $N_{\tau}=a^{3}(\tau)$, and the the Hamiltonian
\eqref{Hamiltonian-SF1} of the test field reduces to 
\begin{equation}
H_{\varphi}^{{\rm (iso)}}=\sum_{\mathbf{k}}H_{\tau,\mathbf{k}}\coloneqq\frac{1}{2}\sum_{\mathbf{k}}\Big(\mathsf{p}_{\mathbf{k}}^{2}+\omega_{\tau,k}^{2}\mathsf{q}_{\mathbf{k}}^{2}\Big).\label{Hamiltonian-SF-FLRW}
\end{equation}
Here, the time-dependent frequency for each mode, is obtained by substituting
$a_{1}=a_{2}=a_{3}\equiv a(\tau)$ in Eq. \eqref{frequency-SF-b1},
as $\omega_{\tau,k}^{2}\equiv k^{2}a^{4}+m^{2}a^{6}$. Like before,
the background elements consist of the spacetime plus a background
matter source in form of a \emph{massless} scalar field $\phi(\tau)$
which serves as an internal physical time parameter \cite{Ashtekar:2006rx}.
The propagating test matter field is denoted by the scalar field $\varphi$.
In this case, not only $\varphi$ but also the background spacetime
is to be quantized. This will lead to a theory of a quantum test field
$\varphi$, whose wave function, denoted by $\psi$, evolves with
respect to the internal time variable $\phi$ on the background quantum
geometry. Due to neglecting backreaction, for a given mode $\mathbf{k}$,
the full kinematical Hilbert space of the system is given by ${\cal H}_{{\rm kin}}^{(\mathbf{k})}={\cal H}_{{\rm kin}}^{o}\otimes{\cal H}_{\varphi}^{(\mathbf{k})}$,
where ${\cal H}_{{\rm kin}}^{o}={\cal H}_{{\rm grav}}\otimes{\cal H}_{\phi}$
is the background Hilbert space consisting of the Hilbert space of
the geometry and the scalar clock variable $\phi$ (again with no
backreaction). The matter sectors are quantized using the Schrödinger
representation, with the Hilbert spaces ${\cal H}_{\varphi}^{(\mathbf{k})}=L_{2}(\mathbb{R},d\mathsf{q}_{\mathbf{k}})$
and ${\cal H}_{\phi}=L_{2}(\mathbb{R},d\phi)$. For any physical state
$\Psi(\nu,\mathsf{q}_{\mathbf{k}},\phi)\in{\cal H}_{{\rm kin}}^{(\mathbf{k})}$,
with $\nu$ the quantum number related to the geometry (see below),
the action of the full Hamiltonian constraint operator $\hat{{\cal C}}_{\tau,\mathbf{k}}$
for the $\mathbf{k}$'th mode is written as \cite{Ashtekar:2009mb}
\begin{equation}
\hat{{\cal C}}_{\tau,\mathbf{k}}\Psi=\big(N_{\tau}\hat{{\cal C}}_{o}+\hat{H}_{\tau,\mathbf{k}}\big)\Psi=0,\label{const-tot}
\end{equation}
where $\hat{{\cal C}}_{o}=\hat{{\cal C}}_{{\rm grav}}+\hat{{\cal C}}_{\phi}$
is the background scalar constraint operator and 
\begin{equation}
\hat{H}_{\tau,\mathbf{k}}=\frac{1}{2}\left[\hat{\mathsf{p}}_{\mathbf{k}}^{2}+\left(k^{2}\hat{a}^{4}+m^{2}\hat{a}^{6}\right)\hat{\mathsf{q}}_{\mathbf{k}}^{2}\right],\label{eq:H-tau-k}
\end{equation}
is the Hamiltonian of the test field $\varphi$. The background term
$\hat{{\cal C}}_{o}$ is well-defined on ${\cal H}_{{\rm kin}}^{o}$,
so that, the physical states $\Psi_{o}(\phi,\nu)\in{\cal H}_{{\rm kin}}^{o}$
are those lying on the kernel of $\hat{{\cal C}}_{o}$, and are solutions
to a self-adjoint Hamiltonian constraint equation of the form \cite{Ashtekar:2006rx,Ashtekar:2006uz,Ashtekar:2006wn}
\begin{equation}
N_{\tau}\hat{{\cal C}}_{o}\Psi_{o}(\nu,\phi)=-\frac{\hbar^{2}}{2\ell^{3}}(\partial_{\phi}^{2}+\Theta)\Psi_{o}(\nu,\phi)=0,\label{H-constraint1}
\end{equation}
where $\Theta$ is a difference operator that acts on $\Psi_{o}$
and involves only the gravitational sector $\nu$ but not $\phi$.
The quantum number $\nu$ is the eigenvalue of the volume operator
of the isotropic background geometry $\hat{V}_{o}=\widehat{\ell^{3}a^{3}}$,
which acts on $\Psi_{o}$ as $\hat{V}_{o}\Psi_{o}(\nu,\phi)=2\pi\gamma\ell_{{\rm Pl}}|\nu|\Psi_{o}(\nu,\phi)$.

By restricting to the space spanned by the positive frequency solutions
to Eq. \eqref{H-constraint1}, one can write a Schrödinger equation
for the background sector \cite{Ashtekar:2006rx} 
\begin{equation}
-i\hbar\partial_{\phi}\Psi_{o}(\nu,\phi)=\hbar\sqrt{\Theta}\Psi_{o}(\nu,\phi)\eqqcolon\hat{H}_{o}\Psi_{o}(\nu,\phi).
\end{equation}
This constructs the physical Hilbert space ${\cal H}_{{\rm phys}}^{o}$
of the geometry, endowed with the scalar product 
\begin{equation}
\langle\Psi_{o}|\Psi_{o}^{\prime}\rangle=\sum_{\nu}\Psi_{o}^{\ast}(\nu,\phi_{0})\Psi_{o}^{\prime}(\nu,\phi_{0}),
\end{equation}
for any ``instant'' of internal time $\phi_{0}$. By substituting
Eqs. \eqref{H-constraint1} and \eqref{eq:H-tau-k} in Eq. \eqref{const-tot},
one obtains a Schrödinger equation for the full state $\Psi\left(\nu,\mathsf{q}_{\mathbf{k}},\phi\right)$
of the background-test field system as \cite{Ashtekar:2009mb} 
\begin{equation}
-i\hbar\partial_{\phi}\Psi(\nu,\mathsf{q}_{\mathbf{k}},\phi)\approx\big(\hat{H}_{o}-\hat{H}_{\phi,\mathbf{k}}\big)\Psi(\nu,\mathsf{q}_{\mathbf{k}},\phi),\label{shro-eq2-approx}
\end{equation}
with
\begin{equation}
\hat{H}_{\phi,\mathbf{k}}\coloneqq\ell^{3}\hat{H}_{o}^{-\frac{1}{2}}\hat{H}_{\tau,\mathbf{k}}\hat{H}_{o}^{-\frac{1}{2}}.
\end{equation}
In deriving Eq.~\eqref{shro-eq2-approx}, we have used a test field
approximation, by disregarding the backreaction of the matter on the
background homogeneous geometry, so that, the term $\hat{H}_{\tau,\mathbf{k}}$
can be seen as the Hamiltonian of a small perturbation added to the
geometrical sector $\hat{H}_{o}^{2}$. It should be noted that, in
the present formalism, $\phi$ is well-suited to be considered an
emergent time in quantum theory. So, Eq. \eqref{shro-eq2-approx}
indicates a quantum evolution of the state $\Psi(\nu,\mathsf{q}_{\mathbf{k}},\phi)$
with respect to the internal physical time $\phi$, which depends
on the $\mathbf{k}$'th mode of the test field $\varphi$, and the
quantum geometry encoded in $\nu$.

To find the evolution equation of the wave function $\psi$ of the
test field on the this quantum FLRW background spacetime, we
work in an interaction picture. This is achieved by introducing $\Psi_{{\rm int}}(\nu,\mathsf{q}_{\mathbf{k}},\phi)=\exp[-(i\hat{H}_{o}/\hbar)(\phi-\phi_{0})]\Psi(\nu,\mathsf{q}_{\mathbf{k}},\phi)$
in Eq. \eqref{shro-eq2-approx}, which regards $\hat{H}_{o}$ as the
Hamiltonian of the heavy degree of freedom, and $\hat{H}_{\phi,\mathbf{k}}$
as the Hamiltonian of the light degree of freedom (i.e., a perturbation
term). The relational time $\phi_{0}$ is, as before, any fixed instant
of time. As mentioned before, by disregarding the backreaction of
the test field on geometry we can approximate $\Psi(\nu,\mathsf{q}_{\mathbf{k}},\phi)=\Psi_{o}(\nu,\phi)\otimes\psi(\mathsf{q}_{\mathbf{k}},\phi)$.
This implies that the geometry evolves by Hamiltonian $\hat{H}_{o}$
as $\hat{p}_{\phi}\Psi_{o}(\nu,\phi)=-i\hbar\partial_{\phi}\Psi_{o}(\nu,\phi)=\hat{H}_{o}\Psi_{o}(\nu,\phi)$
for any $\Psi_{o}\in{\cal H}_{{\rm kin}}^{o}$ in the Heisenberg picture,
thus, we can write $\Psi_{o}(\nu,\phi)=\exp[(i\hat{H}_{o}/\hbar)(\phi-\phi_{0})]\Psi_{o}(\nu,\phi_{0})$.
This leads to the factorization of $\Psi_{{\rm int}}$ as $\Psi_{{\rm int}}(\nu,\mathsf{q}_{\mathbf{k}},\phi)=\Psi_{o}(\nu,\phi_{0})\otimes\psi(\mathsf{q}_{\mathbf{k}},\phi)$.
Now, plugging $\Psi_{{\rm int}}$ into \eqref{shro-eq2-approx} and
projecting both sides on $\Psi_{o}(\nu,\phi_{0})$ yields an evolution
equation for $\psi(\mathsf{q}_{\mathbf{k}},\phi)$ as \cite{Ashtekar:2009mb}
\begin{align}
i\hbar\partial_{\phi}\psi\left(\mathsf{q}_{\mathbf{k}},\phi\right)= &\ \frac{\ell^{3}}{2}\left[\langle\hat{H}_{o}^{-1}\rangle\hat{\mathsf{p}}_{\mathbf{k}}^{2}\right.\nonumber \\
 & \left.+\left(\sum_{i}k_{i}^{2}\left\langle \hat{H}_{o}^{-\frac{1}{2}}\hat{a}^{4}(\phi)\hat{H}_{o}^{-\frac{1}{2}}\right\rangle +m^{2}\left\langle \hat{H}_{o}^{-\frac{1}{2}}\hat{a}^{6}(\phi)\hat{H}_{o}^{-\frac{1}{2}}\right\rangle \right)\hat{\mathsf{q}}_{\mathbf{k}}^{2}\right]\psi\left(\mathsf{q}_{\mathbf{k}},\phi\right)\label{evol-eq1}
\end{align}
where $\left\langle \cdot\right\rangle $ denote the expectation value
with respect to the quantum state $\Psi_{o}(\nu,\phi_{0})$. In this
interaction picture, the state $\Psi_{o}(\nu,\phi_{0})$ of the quantum
geometry is described in Heisenberg picture which is frozen at time
$\phi=\phi_{0}$, while the geometrical operator $\hat{a}(\phi)=\hat{V}_{o}^{1/3}(\phi)/\ell$
evolves in time as $\hat{a}(\phi)=\exp[-(i\hat{H}_{o}/\hbar)(\phi-\phi_{0})]\hat{a}\exp[(i\hat{H}_{o}/\hbar)(\phi-\phi_{0})]$.
In addition, the state of the test field $\psi(\mathsf{q}_{\mathbf{k}},\phi)$,
unlike $\Psi_{o}$, evolves in time $\phi$ due to the evolution equation
\eqref{evol-eq1} in Schrödinger picture.

\subsection{Emergence of effective anisotropies from quantum FLRW background\label{solution-Bianchi}}

A comparison between, and matching of the evolution equations \eqref{evol-eq1}
and \eqref{Hamiltonian-SF-bquantum-eff} reveals that due to the quantum
effects, the original FLRW metric can be interpreted to be now replaced
by an effective anisotropic geometry, with the lapse function $\tilde{N}_{\phi}$,
scale factors $\tilde{a}_{i}$ and the momentum $\tilde{p}_{\phi}$,
which acts as the background for propagation of the quantum modes.
In other words, the quantum FLRW resembles an emergent anisotropic
classical Bianchi I model with tilde variables. Note that we have
considered the general case where both mass $\tilde{m}$ and the wave
vector $\tilde{k}$ in \eqref{Hamiltonian-SF-bquantum-eff} can be
different (or dressed) compared to the original FLRW system, as a
consequence of the quantum spacetime effects. The relations between
variables of this effective dressed geometry (classical Bianchi I
with tilde) and those of the isotropic quantum (FLRW) geometry are
seen to be 
\begin{align}
\tilde{N}_{\phi}= &\ \ell^{3}\left|\tilde{a}_{1}\tilde{a}_{2}\tilde{a}_{3}\right|\left\langle \hat{H}_{o}^{-1}\right\rangle ,\label{eq:main-sys-1}\\
\sum_{i=1}^{3}\left(\frac{\tilde{k}_{i}}{\tilde{a}_{i}}\right)^{2}\tilde{N}_{\phi}\frac{\left(\tilde{a}_{1}\tilde{a}_{2}\tilde{a}_{3}\right)^{2}}{\left|\tilde{a}_{1}\tilde{a}_{2}\tilde{a}_{3}\right|}= &\ \sum_{i=1}^{3}k_{i}^{2}\ell^{3}\left\langle \hat{H}_{o}^{-\frac{1}{2}}\hat{a}^{4}\hat{H}_{o}^{-\frac{1}{2}}\right\rangle ,\label{eq:main-sys-2}\\
\tilde{N}_{\phi}\tilde{m}^{2}\frac{\left(\tilde{a}_{1}\tilde{a}_{2}\tilde{a}_{3}\right)^{2}}{\left|\tilde{a}_{1}\tilde{a}_{2}\tilde{a}_{3}\right|}= &\ \ell^{3}m^{2}\left\langle \hat{H}_{o}^{-\frac{1}{2}}\hat{a}^{6}\hat{H}_{o}^{-\frac{1}{2}}\right\rangle ,\label{eq:main-sys-3}
\end{align}
in which $\hat{a}=\hat{a}(\phi)$, and $\langle\hat{H}_{o}^{-1}\rangle=(\tilde{p}_{\phi})^{-1}$.
It is seen that we have a system of five equations \eqref{eq:main-sys-1},
\eqref{eq:main-sys-2} and \eqref{eq:main-sys-3} (that is, if we
only identify the terms with $\tilde{k}_{i}$ and $k_{i}$ with each
other) with eight unknowns (the tilde variables), which constitutes
an underdetermined system. To be able to solve the system we need
three more equations (i.e. conditions) on these variables. These conditions
can be imposed on either of $\tilde{m}\,,\tilde{k}_{i},\,\tilde{a}_{i}$
or $\tilde{N}_{\phi}$. In what follows, we will investigate some
of these conditions and their physical consequences on the solutions
by dividing the discussion into three cases: massless test field $m=0=\tilde{m}$,
undressed mass $\tilde{m}=m$, or dressed mass $\tilde{m}\neq m$.

\subsubsection{Massless case ($m=0=\tilde{m}$)}

If $m=0$, we will immediately obtain $\tilde{m}=0$. But the situation
is not changed and we still need three more conditions to be able
to solve the system above, since we have also eliminated one of our
equations. The simplest option is to take the conditions $k_{i}^{2}=\tilde{k}_{i}^{2}$
(for each $i$). Then we will get $\tilde{a}_{1}^{2}=\tilde{a}_{2}^{2}=\tilde{a}_{3}^{2}=\tilde{a}^{2}$
and from this one obtains
\begin{align}
\tilde{a}^{4}= &\ \frac{\left\langle \hat{H}_{o}^{-\frac{1}{2}}\hat{a}^{4}(\phi)\hat{H}_{o}^{-\frac{1}{2}}\right\rangle }{\left\langle \hat{H}_{o}^{-1}\right\rangle },\label{massless-dressed-scale-1}\\
\tilde{N}_{\phi}= &\ \ell^{3}\left\langle \hat{H}_{o}^{-1}\right\rangle ^{\frac{1}{4}}\left\langle \hat{H}_{o}^{-\frac{1}{2}}\hat{a}^{4}(\phi)\hat{H}_{o}^{-\frac{1}{2}}\right\rangle ^{\frac{3}{4}}=\ell^{3}\tilde{p}_{\phi}^{-1}\tilde{a}^{3}.\label{massless-dressed-scale-2}
\end{align}
A same result is yielded if, instead, one takes the assumption that
the underlying dressed geometry is an effective FLRW metric, i.e.,
$\tilde{a}_{1}=\tilde{a}_{2}=\tilde{a}_{3}=\tilde{a}(\phi)$ (two
conditions) and that one of the components of $\tilde{\mathbf{k}}$
is equal to the corresponding component of the undressed $\mathbf{k}$
\cite{Ashtekar:2009mb}.

\subsubsection{Undressed mass ($\tilde{m}=m\protect\neq0$)}

Since having an undressed mass gives us one of the three additional
conditions needed, we are left with two conditions on either $\tilde{k}_{i}$
or $\tilde{a}_{i}$ (forgetting about $\tilde{N}_{\phi}$ since its
conditions can be found from those of $\tilde{a}_{i}$). In what follows,
we express the relations between $\tilde{k}_{i},\tilde{a}_{i}$, etc.,
using some arbitrary coefficients $\alpha_{ij,}\,\beta_{ij},\,\gamma_{ij},\,\rho_{ij}$
and $\sigma_{ij}$. \\

\noindent \textbf{Case 1:} Let us first impose the two remaining conditions
just on $\tilde{k}_{i}$, and between the tilde variables themselves.
In general we can have
\begin{equation}
\tilde{k}_{1}=\alpha_{12}\tilde{k}_{2},\,\,\,\,\,\,\tilde{k}_{2}=\alpha_{23}\tilde{k}_{3}.
\end{equation}
If we also naturally define the relation between $\tilde{k}_{1}$
and $\tilde{k}_{3}$ as $\tilde{k}_{1}=\alpha_{13}\tilde{k}_{3}$,
and, furthermore, define the inverse relations $\tilde{k}_{2}=\alpha_{21}\tilde{k}_{1},\,\tilde{k}_{3}=\alpha_{32}\tilde{k}_{2},\,\tilde{k}_{3}=\alpha_{31}\tilde{k}_{1}$,
we obtain 
\begin{align}
\alpha_{ij}= &\ \frac{1}{\alpha_{ji}},\,\,\,\,\,\,\,(i,j,k=1,2,3).\label{eq:alpha-rel-1}\\
\alpha_{ij}= &\ \alpha_{ik}\alpha_{kj},\,\,\,\,\,\,\,(i,j,k=1,2,3;\,i\neq j,i\neq k,j\neq k),\label{eq:alpha-rel-2}
\end{align}
These can then be used to write equations \eqref{eq:main-sys-1}-\eqref{eq:main-sys-3}
as
\begin{align}
\tilde{N}_{\phi}= &\ \ell^{3}\left\langle \hat{H}_{o}^{-1}\right\rangle ^{\frac{1}{2}}\left\langle \hat{H}_{o}^{-\frac{1}{2}}\hat{a}^{6}(\phi)\hat{H}_{o}^{-\frac{1}{2}}\right\rangle ^{\frac{1}{2}},\label{eq:rainbw-1}\\
\tilde{a}_{i}^{6}= &\ \frac{1}{2}\sum_{j,k=1}^{3}\left|\epsilon_{ijk}\right|\alpha_{ij}^{2}\alpha_{ik}^{2}\frac{k_{j}^{2}k_{k}^{2}}{k_{i}^{4}}\frac{\left\langle \hat{H}_{o}^{-\frac{1}{2}}\hat{a}^{6}(\phi)\hat{H}_{o}^{-\frac{1}{2}}\right\rangle }{\left\langle \hat{H}_{o}^{-1}\right\rangle },\label{eq:rainbw-2}\\
\tilde{k}_{i}^{6}= &\ \frac{1}{2}\sum_{j,k=1}^{3}\left|\epsilon_{ijk}\right|\alpha_{ij}^{2}\alpha_{ik}^{2}k_{i}^{2}k_{j}^{2}k_{k}^{2}\frac{\left\langle \hat{H}_{o}^{-\frac{1}{2}}\hat{a}^{4}(\phi)\hat{H}_{o}^{-\frac{1}{2}}\right\rangle ^{3}}{\left\langle \hat{H}_{o}^{-\frac{1}{2}}\hat{a}^{6}(\phi)\hat{H}_{o}^{-\frac{1}{2}}\right\rangle ^{2}\left\langle \hat{H}_{o}^{-1}\right\rangle },\label{eq:rainbw-3}
\end{align}
where $i=1,2,3$. The above equations may seem like a {\em rainbow metric}, given that the components of the dressed metric are now mode  dependent (cf. e.g. \cite{Assaniousssi:2014ota,Magueijo:2002xx}). In such a case, the modified scale factors and consequently the metric, acquire a dependence on the wave vectors. Thereby, the scale factor that the propagating wave	``sees'' in each direction, depends on its wave vector $\mathbf{k}$ in that direction. But, it should be noted that, the (dressed) wave vector associated to the dressed scale factors $\tilde{a}$ is the $\tilde{\mathbf{k}}$ and not $\mathbf{k}$. Thus, if one writes the above equations in terms of $\tilde{a}_{i}$ and the {\em dressed} mode components $\tilde{k}_i$, one finds out that this is not the case. Nevertheless, these equations offer a detailed insight into the dependence of the unmodified wave vector components and the modified scale factors.\\

\noindent \textbf{Case 2:} Let us impose the conditions between only
two components of $\tilde{k}_{i}$ and $k_{i}$. Without loss of generality,
we can assume $i=1,2$ and hence
\begin{equation}
\tilde{k}_{1}=\beta_{11}k_{1},\,\,\,\,\,\,\tilde{k}_{2}=\beta_{22}k_{2}.\label{k1k2k3-1}
\end{equation}
Then we will have
\begin{align}
\tilde{N}_{\phi}= &\ \ell^{3}\left\langle \hat{H}_{o}^{-1}\right\rangle ^{\frac{1}{2}}\left\langle \hat{H}_{o}^{-\frac{1}{2}}\hat{a}^{6}(\phi)\hat{H}_{o}^{-\frac{1}{2}}\right\rangle ^{\frac{1}{2}},\label{k1k2k3-2}\\
\tilde{a}_{i}^{2}= &\ \beta_{ii}^{2}\frac{\left\langle \hat{H}_{o}^{-\frac{1}{2}}\hat{a}^{6}(\phi)\hat{H}_{o}^{-\frac{1}{2}}\right\rangle }{\left\langle \hat{H}_{o}^{-\frac{1}{2}}\hat{a}^{4}(\phi)\hat{H}_{o}^{-\frac{1}{2}}\right\rangle },\,\,\,\,\,\,\,\,\,\,\,\,\,\,\,(i=1,2)\label{k1k2k3-3}\\
\tilde{a}_{3}^{2}= &\ \frac{1}{\beta_{22}^{2}\beta_{11}^{2}}\frac{\left\langle \hat{H}_{o}^{-\frac{1}{2}}\hat{a}^{4}(\phi)\hat{H}_{o}^{-\frac{1}{2}}\right\rangle ^{2}}{\left\langle \hat{H}_{o}^{-\frac{1}{2}}\hat{a}^{6}(\phi)\hat{H}_{o}^{-\frac{1}{2}}\right\rangle \left\langle \hat{H}_{o}^{-1}\right\rangle },\label{k1k2k3-4}\\
\tilde{k}_{3}^{2}= &\ \frac{k_{3}^{2}}{\beta_{11}^{2}\beta_{22}^{2}}\frac{\left\langle \hat{H}_{o}^{-\frac{1}{2}}\hat{a}^{4}(\phi)\hat{H}_{o}^{-\frac{1}{2}}\right\rangle ^{3}}{\left\langle \hat{H}_{o}^{-1}\right\rangle \left\langle \hat{H}_{o}^{-\frac{1}{2}}\hat{a}^{6}(\phi)\hat{H}_{o}^{-\frac{1}{2}}\right\rangle ^{2}}.\label{k1k2k3-5}
\end{align}
As can be seen from the scale factors above, the emergent dressed
spacetime becomes anisotropic. Hence in this case, we have an effective
theory in which a field with a dressed wave vector $\tilde{\mathbf{k}}$
propagates on a semiclassical dressed spacetime whose isotropy is
broken. Furthermore, the form of this isotropy breaking depends on
the components of the wave vector $\tilde{\mathbf{k}}$, which itself
depends on the form of dependence of $\tilde{\mathbf{k}}$ on $\mathbf{k}$
in \eqref{k1k2k3-1}.\\

\noindent \textbf{Case 3:} Here we put the conditions just on $\tilde{a}_{i}$,
and again between the tilde variables themselves. Thus, in general
we can have
\begin{equation}
\tilde{a}_{1}=\gamma_{12}\tilde{a}_{2},\,\,\,\,\,\,\,\,\,\,\tilde{a}_{2}=\gamma_{23}\tilde{a}_{3},\label{a1a213-1}
\end{equation}
which, using \eqref{eq:alpha-rel-1} and \eqref{eq:alpha-rel-2},
leads to
\begin{align}
\tilde{N}_{\phi}= &\ \ell^{3}\left\langle \hat{H}_{o}^{-1}\right\rangle ^{\frac{1}{2}}\left\langle \hat{H}_{o}^{-\frac{1}{2}}\hat{a}^{6}(\phi)\hat{H}_{o}^{-\frac{1}{2}}\right\rangle ^{\frac{1}{2}},\label{a1a213-2}\\
\tilde{a}_{i}^{6}= &\ \frac{1}{2}\sum_{j,k=1}^{3}\left|\epsilon_{ijk}\right|\gamma_{ij}^{2}\gamma_{ik}^{2}\frac{\left\langle \hat{H}_{o}^{-\frac{1}{2}}\hat{a}^{6}(\phi)\hat{H}_{o}^{-\frac{1}{2}}\right\rangle }{\left\langle \hat{H}_{o}^{-1}\right\rangle },\label{a1a213-3}\\
\tilde{k}_{i}^{3}= &\ \frac{1}{2}k_{i}^{3}\sum_{j,k=1}^{3}\left|\epsilon_{ijk}\right|\gamma_{ij}\gamma_{ik}\frac{\left\langle \hat{H}_{o}^{-\frac{1}{2}}\hat{a}^{4}(\phi)\hat{H}_{o}^{-\frac{1}{2}}\right\rangle ^{\frac{3}{2}}}{\left\langle \hat{H}_{o}^{-1}\right\rangle ^{\frac{1}{2}}\left\langle \hat{H}_{o}^{-\frac{1}{2}}\hat{a}^{6}(\phi)\hat{H}_{o}^{-\frac{1}{2}}\right\rangle },\label{a1a213-4}
\end{align}
with $i=1,2,3$. These solutions correspond to an effective theory
in which the quantum field propagates on an anisotropic dressed background
with $\tilde{a}_{i}$'s given by \eqref{a1a213-2}. It is seen that
such as field acquires an effective wave vector $\tilde{\mathbf{k}}$
whose components $\tilde{k}_{i}$ are modified with respect to the
original $k$, due to quantum fluctuations of the original isotropic
background.\\

\noindent \textbf{Case 4:} Another type of mixed condition could be
such that we have one relation between $\tilde{a}_{i}$ and $\tilde{a}_{j}$,
and one between $\tilde{k}_{i}$ and $k_{i}$, of which Ref.~\cite{Dapor:2012qn}
is an example. In this case we can, e.g., have
\begin{equation}
\tilde{a}_{1}=\sigma_{12}\tilde{a}_{2},\,\,\,\,\,\,\,\,\,\,\tilde{k}_{3}=\sigma_{33}k_{3}.\label{a1k1-1}
\end{equation}
Using \eqref{eq:alpha-rel-1} and, this leads to
\begin{align}
\tilde{N}_{\phi}= &\ \ell^{3}\left\langle \hat{H}_{o}^{-1}\right\rangle ^{\frac{1}{2}}\left\langle \hat{H}_{o}^{-\frac{1}{2}}\hat{a}^{6}(\phi)\hat{H}_{o}^{-\frac{1}{2}}\right\rangle ^{\frac{1}{2}},\label{a1k1-2}\\
\tilde{a}_{i}^{4}= &\ \left(\frac{\sigma_{il}}{\sigma_{33}}\right)^{2}\frac{\left\langle \hat{H}_{o}^{-\frac{1}{2}}\hat{a}^{4}(\phi)\hat{H}_{o}^{-\frac{1}{2}}\right\rangle }{\left\langle \hat{H}_{o}^{-1}\right\rangle },\,\,\,\,\,\,\,\,\,\,(i,l=1,2;\,l\neq i)\label{a1k1-3}\\
\tilde{a}_{3}^{4}= &\ \sigma_{33}^{4}\frac{\left\langle \hat{H}_{o}^{-\frac{1}{2}}\hat{a}^{6}(\phi)\hat{H}_{o}^{-\frac{1}{2}}\right\rangle ^{2}}{\left\langle \hat{H}_{o}^{-\frac{1}{2}}\hat{a}^{4}(\phi)\hat{H}_{o}^{-\frac{1}{2}}\right\rangle ^{2}},\label{a1k1-4}\\
\tilde{k}_{i}^{2}= &\ k_{i}^{2}\left(\frac{\sigma_{il}}{\sigma_{33}}\right)^{2}\frac{\left\langle \hat{H}_{o}^{-\frac{1}{2}}\hat{a}^{4}(\phi)\hat{H}_{o}^{-\frac{1}{2}}\right\rangle ^{\frac{3}{2}}}{\left\langle \hat{H}_{o}^{-\frac{1}{2}}\hat{a}^{6}(\phi)\hat{H}_{o}^{-\frac{1}{2}}\right\rangle \left\langle \hat{H}_{o}^{-1}\right\rangle ^{\frac{1}{2}}},\,\,\,\,\,\,\,\,\,\,(i,l=1,2;\,l\neq i)\label{a1k1-5}
\end{align}
where as in \eqref{eq:alpha-rel-1}, we have naturally defined $\sigma_{ij}=\sigma_{ji}^{-1}$.
Here again we have a field with an effective $\tilde{\mathbf{k}}$,
propagating on an anisotropic dressed background. However, as can
be seen, one of the direction of $\tilde{k}$ (here $\tilde{k}_{3}$)
plays a rather more special role as opposed to the situation in Case
3.

\noindent Note that in all the above four cases, in the presence of
the nonvanishing undressed mass, the lapse functions $\tilde{N}_{\phi}$
are identical.

\subsubsection{Dressed mass ($\tilde{m}\protect\neq m;\,m,\tilde{m}\protect\neq0$)\label{dressedmass-sol}}

The situation here is a bit different. Since we have removed the condition
on mass, we now need three conditions on $\tilde{a}_{i}$ and $\tilde{k}_{i}$.
Let us consider some of the possible cases below.\\

\noindent \textbf{Case 1:} We first impose these remaining conditions
just on $\tilde{k}_{i}$,
\begin{equation}
\tilde{k}_{i}=\alpha_{ii}k_{i},\,\,\,\,\,\,\left(i=1,2,3\right).
\end{equation}
This will lead to
\begin{align}
\tilde{N}_{\phi}= &\ \frac{\ell^{3}}{\left(\alpha_{11}\alpha_{22}\alpha_{33}\right)^{\frac{1}{2}}}\left\langle \hat{H}_{o}^{-\frac{1}{2}}\hat{a}^{4}(\phi)\hat{H}_{o}^{-\frac{1}{2}}\right\rangle ^{\frac{3}{4}}\left\langle \hat{H}_{o}^{-1}\right\rangle ^{\frac{1}{4}},\label{lapse-massdressed1}\\
\tilde{a}_{i}^{4}= &\ \left(\frac{\alpha_{ii}}{\alpha_{jj}\alpha_{kk}}\right)^{2}\frac{\left\langle \hat{H}_{o}^{-\frac{1}{2}}\hat{a}^{4}(\phi)\hat{H}_{o}^{-\frac{1}{2}}\right\rangle }{\left\langle \hat{H}_{o}^{-1}\right\rangle },\,\,\,\,\,\,\,\,\,(i,j,k=1,2,3;\,i\neq j\neq k)\label{anisotropy-scale1}\\
\tilde{m}^{4}= &\ m^{4}\left(\alpha_{11}\alpha_{22}\alpha_{33}\right)^{2}\frac{\left\langle \hat{H}_{o}^{-\frac{1}{2}}\hat{a}^{6}(\phi)\hat{H}_{o}^{-\frac{1}{2}}\right\rangle ^{2}\left\langle \hat{H}_{o}^{-1}\right\rangle }{\left\langle \hat{H}_{o}^{-\frac{1}{2}}\hat{a}^{4}(\phi)\hat{H}_{o}^{-\frac{1}{2}}\right\rangle ^{3}}.
\end{align}
Note that as a special subcase, when $\alpha_{ii}=1$, one obtains
an undressed wave vector $\tilde{k}_{i}=k_{i}$, a dressed background
with a scale factor $\tilde{a}_{1}=\tilde{a}_{2}=\tilde{a}_{3}=\tilde{a}$
identical to \eqref{massless-dressed-scale-1}
\begin{equation}
\tilde{a}^{4}\left(\phi\right)=\frac{\left\langle \hat{H}_{o}^{-\frac{1}{2}}\hat{a}^{4}(\phi)\hat{H}_{o}^{-\frac{1}{2}}\right\rangle }{\left\langle \hat{H}_{o}^{-1}\right\rangle },\label{anisotropy-scale1b}
\end{equation}
and a dressed mass
\begin{equation}
\tilde{m}^{2}=m^{2}\frac{\left\langle \hat{H}_{o}^{-\frac{1}{2}}\hat{a}^{6}(\phi)\hat{H}_{o}^{-\frac{1}{2}}\right\rangle \left\langle \hat{H}_{o}^{-1}\right\rangle ^{\frac{1}{2}}}{\left\langle \hat{H}_{o}^{-\frac{1}{2}}\hat{a}^{4}(\phi)\hat{H}_{o}^{-\frac{1}{2}}\right\rangle ^{\frac{3}{2}}}.\label{dressed-mass0}
\end{equation}
This corresponds to an isotropic dressed background spacetime with
the scale factor \eqref{massless-dressed-scale-1} over which a test
field with an undressed wave-vector $\mathbf{k}=(k_{1},k_{2},k_{3})$
propagates. Moreover, the mass in this case is now dressed as in \eqref{dressed-mass0}.

A special subcase for this case is when the field modes propagate
only along the $z$-direction, i.e., $k_{1}=k_{2}=0$ and $k=k_{3}$,
and hence, $\tilde{k}_{1}=\tilde{k}_{2}=0$ and $\tilde{k}=\tilde{k}_{3}=\alpha_{33}k$.
Then, by setting $\alpha_{11}=\alpha_{22}=1$, one obtains a dressed
background with an emergent anisotropy as a result of the scale factor
$\tilde{a}_{3}$. In that case, $\tilde{a}_{3}=\tilde{a}_{1}/\alpha_{33}=\tilde{a}_{2}/\alpha_{33}$.
Similar argument was presented in Ref. \cite{Dapor:2012qn}. However,
in our case, in addition to the scenario presented in \cite{Dapor:2012qn},
the anisotropy can still exist even if the field is massless.\\

\noindent \textbf{Case 2:} Another choice is to impose two conditions
between $\tilde{a}_{i}$, and one between $\tilde{k}_{i}$ and $k_{i}$,
as, e.g.,
\begin{equation}
\tilde{a}_{1}=\rho_{12}\tilde{a}_{2},\quad\tilde{a}_{2}=\rho_{23}\tilde{a}_{3},\quad\tilde{k}_{3}=\rho_{33}k_{3}\label{a1a2k3-1}
\end{equation}
This way one gets
\begin{align}
\tilde{N}_{\phi}= &\ \frac{\ell^{3}}{\rho_{12}^{\frac{1}{2}}\rho_{23}\rho_{33}^{\frac{3}{2}}}\left\langle \hat{H}_{o}^{-\frac{1}{2}}\hat{a}^{4}(\phi)\hat{H}_{o}^{-\frac{1}{2}}\right\rangle ^{\frac{3}{4}}\left\langle \hat{H}_{o}^{-1}\right\rangle ^{\frac{1}{4}},\label{lapse-massdressed2}\\
\tilde{a}_{3}^{4}= &\ \frac{1}{\rho_{33}^{2}\rho_{12}^{2}\rho_{23}^{4}}\frac{\left\langle \hat{H}_{o}^{-\frac{1}{2}}\hat{a}^{4}(\phi)\hat{H}_{o}^{-\frac{1}{2}}\right\rangle }{\left\langle \hat{H}_{o}^{-1}\right\rangle },\label{a1a2k3-3}\\
\tilde{k}_{1}^{2}= &\ \left(\rho_{12}\rho_{23}\rho_{33}\right)^{2}k_{1}^{2},\\
\tilde{k}_{2}^{2}= &\ \left(\rho_{23}\rho_{33}\right)^{2}k_{2}^{2},\\
\tilde{m}^{2}= &\  m^{2}\rho_{12}\rho_{23}^{2}\rho_{33}^{3}\frac{\left\langle \hat{H}_{o}^{-\frac{1}{2}}\hat{a}^{6}(\phi)\hat{H}_{o}^{-\frac{1}{2}}\right\rangle \left\langle \hat{H}_{o}^{-1}\right\rangle ^{\frac{1}{2}}}{\left\langle \hat{H}_{o}^{-\frac{1}{2}}\hat{a}^{4}(\phi)\hat{H}_{o}^{-\frac{1}{2}}\right\rangle ^{\frac{3}{2}}}.
\end{align}
It is seen that in this case, assuming nonvanishing coefficients
$\rho_{12},\rho_{23},\rho_{33}$ represents the propagation of a quantum
field with a dressed wave vector, over an anisotropic dressed background.
There is a special direction, in this case $\tilde{k}_{3}$, which
is the distinct source of the emergent anisotropy.

\section{Quantum particle production\label{pairprod}}

In this section, we investigate the quantum gravity effects, due to
the emerged anisotropy of the background geometry, on the creation
of particles from Planck regime. Let us rewrite the Schrödinger equation
\eqref{Hamiltonian-SF-bquantum-eff} as 
\begin{equation}
i\hbar\partial_{\phi}\psi=\left(\frac{\hat{\mathsf{p}}_{\mathbf{k}}^{2}}{2M}+\frac{1}{2}M\tilde{\omega}_{k}^{2}(\phi)\,\hat{\mathsf{q}}_{\mathbf{k}}^{2}\right)\psi,\label{Hamiltonian-SF-bquantum-mod}
\end{equation}
where the (internal time) $\phi$-dependent frequency is given by
\begin{equation}
\tilde{\omega}_{k}^{2}(\phi)=\tilde{N}_{\phi}^{2}\left(\sum_{i}^{3}\tilde{k}_{i}^{2}\tilde{a}_{i}^{-2}+\tilde{m}^{2}\right),\label{frequency-1}
\end{equation}
and $M$ is defined as 
\begin{equation}
M=\frac{\left|\tilde{a}_{1}\tilde{a}_{2}\tilde{a}_{3}\right|}{\tilde{N}_{\phi}}=\frac{1}{\ell^{3}\langle\hat{H}_{o}^{-1}\rangle}.\label{eq:M}
\end{equation}
Once the corresponding solutions to $\tilde{N}(\phi)$, $\tilde{k}(\phi)$,
$\tilde{a}_{i}(\phi)$ and $\tilde{m}(\phi)$ are known in terms of
the original quantum geometry fluctuations, the frequency of the each
mode of the field can be computed. It turns out that in all of the
cases of the solutions we discussed in the previous section, despite
the emergence of various types of the dressed geometries (e.g., being
either isotropic or anisotropic), because of the assumption of equality
between Eqs.~\eqref{Hamiltonian-SF-bquantum-eff} and (\ref{evol-eq1}),
the dispersion relation (\ref{frequency-1}) is always of the form
\begin{equation}
\tilde{\omega}_{k}^{2}(\phi)\ =\ \ell^{3}\langle\hat{H}_{o}^{-1}\rangle\left[k^{2}\tilde{a}^{4}+\tilde{m}^{2}\tilde{a}^{6}\right].\label{eq:mod-disp-rel}
\end{equation}
in which $\tilde{a}(\phi)$ is given by  Eq.~(\ref{anisotropy-scale1b})
and the dressed mass $\tilde{m}$ is given by  Eq.~(\ref{dressed-mass0}).
This indicates that, for all possible solutions of the emergent dressed
background, the dispersion relation of the field propagating on it
is modified and takes the same form as in the isotropic dressed background
\eqref{massless-dressed-scale-1}, but with the nonzero dressed mass
\eqref{dressed-mass0}. 

The (classical) equation of motion for each mode $\mathbf{k}$ of
the scalar field, given by the dressed Hamiltonian on the right hand
side of Eq. \eqref{Hamiltonian-SF-bquantum-mod}, is 
\begin{equation}
\ddot{\mathsf{q}}_{\mathbf{k}}+\frac{\dot{M}}{M}\dot{\mathsf{q}}_{\mathbf{k}}+\tilde{\omega}_{k}^{2}\mathsf{q}_{\mathbf{k}}=0,\label{motion1-mod}
\end{equation}
in which a dot denotes a derivative with respect to the internal time
$\phi$. We are interested in studying the particle creation mechanism
using the approach presented in Ref. \cite{Berger:1975ag}. From
Eq.~(\ref{eq:M}) we have that $M=1/(\ell^{3}\langle\hat{H}_{o}^{-1}\rangle)$,
so the term $\dot{M}/M$ in \eqref{motion1-mod} vanishes to a good
approximation if $M$ varies very slowly with time. Although the components
of the dressed emergent metric $\tilde{g}_{ab}$ depend on the fluctuations
of the background quantum geometry operators, it is now a smooth tensor
field so that it is straightforward to pass to a new harmonic time
coordinate $\tilde{\tau}$ (with a lapse function $\tilde{N}_{\tilde{\tau}}=\left|\tilde{a}_{1}\tilde{a}_{2}\tilde{a}_{3}\right|$).
By setting $\tilde{N}_{\tilde{\tau}}d\tilde{\tau}=\tilde{N}_{\phi}d\phi$,
we can simplify our equation of motion \eqref{motion1-mod} through
the time $\tilde{\tau}$: 
\begin{equation}
d\tilde{\tau}=\ell^{3}\langle\hat{H}_{o}^{-1}\rangle d\phi.
\end{equation}
Now, the $\tilde{\tau}$-evolution of the wave function $\psi$ is
given by the reduced Schrödinger equation 
\begin{equation}
i\hbar\partial_{\tau}\psi\left(\tau,\mathsf{q}_{\mathbf{k}}\right)=\frac{1}{2}\left(\hat{\mathsf{p}}_{\mathbf{k}}^{2}+\tilde{\omega}_{k}^{2}(\tau)\hat{\mathsf{q}}_{\mathbf{k}}^{2}\right)\psi\left(\tau,\mathsf{q}_{\mathbf{k}}\right),\label{Hamiltonian-SF-bquantum-mod-newtime}
\end{equation}
where, $\tilde{\omega}_{\tilde{\tau},k}$ now becomes 
\begin{align}
\tilde{\omega}_{\tilde{\tau},k}^{2}\left(\tilde{\tau}\right)= &\ \left(\tilde{a}_{1}\tilde{a}_{2}\tilde{a}_{3}\right)^{2}\left(\sum_{i}^{3}\tilde{k}_{i}^{2}\,\tilde{a}_{i}^{-2}+\tilde{m}^{2}\right)\nonumber \\
= &\  k^{2}\frac{\left\langle \hat{H}_{o}^{-\frac{1}{2}}\hat{a}^{4}\hat{H}_{o}^{-\frac{1}{2}}\right\rangle }{\langle\hat{H}_{o}^{-1}\rangle}+m^{2}\frac{\left\langle \hat{H}_{o}^{-\frac{1}{2}}\hat{a}^{6}\hat{H}_{o}^{-\frac{1}{2}}\right\rangle }{\langle\hat{H}_{o}^{-1}\rangle}.\label{omega-BO}
\end{align}
We assume that the classical equation of motion \eqref{motion1-mod}
has a solution $v_{k}(\tilde{\tau})$, so 
\begin{equation}
v_{k}^{\prime\prime}+\tilde{\omega}_{\tilde{\tau},k}^{2}(\tilde{\tau})v_{k}=0,\label{motion1-mod-harmonic}
\end{equation}
in which a prime denotes a derivative with respect to $\tilde{\tau}$.
Next, we need to find the wave function solutions $\chi_{n}\left(\mathsf{q}_{\mathbf{k}},\tilde{\tau}\right)$
of the differential equation corresponding to \eqref{Hamiltonian-SF-bquantum-mod-newtime}
\begin{equation}
\left(-\frac{\hbar^{2}}{2}\frac{d^{2}}{d\mathsf{q}_{\mathbf{k}}^{2}}+\frac{1}{2}\tilde{\omega}_{\tilde{\tau},k}^{2}\left(\tilde{\tau}\right)\mathsf{q}_{\mathbf{k}}^{2}\right)\chi_{n}\left(\mathsf{q}_{\mathbf{k}},\tilde{\tau}\right)=E_{n}\chi_{n}\left(\mathsf{q}_{\mathbf{k}},\tilde{\tau}\right).
\end{equation}
A complete set of solutions for this differential equation exists
and can be characterized by the quantum number $n$ as \cite{Perelmov:1969,Berger:1975ag}
\begin{equation}
\chi_{n}\left(\mathsf{q}_{\mathbf{k}},\tilde{\tau}\right)=\left(\frac{(v_{k}^{\ast})^{n}}{2^{n}n!\sqrt{2\pi}(v_{k})^{n+1}}\right)^{1/2}\exp\left[\frac{i}{2}\frac{v_{k}^{\prime}}{v_{k}}\mathsf{q}_{\mathbf{k}}^{2}\right]\mathrm{H}_{n}\left(\frac{\mathsf{q}_{\mathbf{k}}}{\sqrt{2}|v_{k}|}\right),\label{number-state}
\end{equation}
with the eigenenergy 
\begin{equation}
E_{n}=\left(n+\frac{1}{2}\right)\hbar\tilde{\omega}_{\tilde{\tau},k},
\end{equation}
where $v_{k}(\tilde{\tau})$ is the solution of Eq. \eqref{motion1-mod-harmonic},
and $\mathrm{H}_{n}$ are the Hermite polynomials of order $n$.

In quantum theory, by introducing the creation and annihilation operators
$\hat{A}_{\mathbf{k}}^{\dagger}$ and $\hat{A}_{\mathbf{k}}$ \cite{Salusti:1970cx}
\begin{equation}
\hat{A}_{\mathbf{k}}=-iv_{k}^{\prime}\left(\tilde{\tau}\right)\mathsf{q}_{\mathbf{k}}+v_{k}\left(\tilde{\tau}\right)\left(\partial/\mathsf{q}_{\mathbf{k}}\right),\label{annihilation}
\end{equation}
where $\hat{A}_{\mathbf{k}}^{\dagger}$ is the Hermitian conjugate
of $\hat{A}_{\mathbf{k}}$, the states \eqref{number-state} can be
generated as 
\begin{equation}
\chi_{n}=\left(n!\right)^{-\frac{1}{2}}\left(\hat{A}_{\mathbf{k}}^{\dagger}\right)^{n}\chi_{0},\quad\quad\quad\hat{A}_{\mathbf{k}}\chi_{0}=0.
\end{equation}
where $[\hat{A}_{\mathbf{k}},\hat{A}_{\mathbf{k}}^{\dagger}]=1$.
Following  \citet{Zeldovich:1971mw}, we assume that there
exists a regime $\tilde{\tau}\leq\tilde{\tau}_{0}$ such that the
vacuum state satisfies an adiabatic condition, so that the usual harmonic
oscillator states may be constructed. We choose the state $|\tilde{0}\rangle$,
satisfying the equation 
\begin{equation}
\frac{1}{2}\tilde{\omega}_{k}\left(\tilde{\tau}_{0}\right)|\tilde{0}\rangle=\hat{H}_{\tilde{\tau},\mathbf{k}}\left(\tilde{\tau}_{0}\right)|\tilde{0}\rangle,
\end{equation}
as the vacuum state, where $\hat{H}_{\tilde{\tau},\mathbf{k}}\left(\tilde{\tau}_{0}\right)$
is the Hamiltonian operator 
\begin{equation}
\hat{H}_{\tilde{\tau},\mathbf{k}}\left(\tilde{\tau}\right)=-\frac{\hbar^{2}}{2}\frac{\partial^{2}}{\partial\mathsf{q}_{\mathbf{k}}^{2}}+\frac{1}{2}\tilde{\omega}_{\tilde{\tau},k}^{2}\left(\tilde{\tau}\right)\hat{\mathsf{q}}_{\mathbf{k}}^{2},\label{Hamiltonian-particle}
\end{equation}
evaluated at some initial time $\tilde{\tau}=\tilde{\tau}_{0}$. Thus,
the state $|\tilde{0}\rangle$ is indeed the harmonic-oscillator ground
state for the frequency $\tilde{\omega}_{\tilde{\tau},k}\left(\tilde{\tau}_{0}\right)$.

Let us now expand the classical test field solution $v_{k}\left(\tilde{\tau}\right)$,
in a Wentzel--Kramers--Brillouin (WKB) approximation, as 
\begin{equation}
v_{k}\left(\tilde{\tau}\right)=\frac{1}{\sqrt{2\tilde{\omega}_{\tilde{\tau},k}}}\exp\left[i\int d\tilde{\tau}\tilde{\omega}_{\tilde{\tau},k}\left(\tilde{\tau}\right)\right].\label{WKB1}
\end{equation}
When the Hamiltonian \eqref{Hamiltonian-particle} possesses an adiabatic
regime, the inequality 
\begin{equation}
\tilde{\omega}_{\tilde{\tau},k}^{\prime}=d\tilde{\omega}_{\tilde{\tau},k}/d\tilde{\tau}\ll\tilde{\omega}_{\tilde{\tau},k}^{2},\label{adiabatic-condtion}
\end{equation}
holds, so that $v_{k}^{\prime}=dv_{k}/d\tilde{\tau}=i\tilde{\omega}_{\tilde{\tau},k}v_{k}$.
Then, the operators $\hat{A}_{\mathbf{k}}$ and $\hat{A}_{\mathbf{k}}^{\dagger}$
reduce to the the usual harmonic oscillator annihilation and creation
operators for the fixed frequency $\tilde{\omega}_{\tilde{\tau},k}$
\begin{align}
\hat{A}_{\mathbf{k}}= &\ v_{k}(\tilde{\tau})\left[\tilde{\omega}_{\tilde{\tau},k}\hat{\mathsf{q}}_{\mathbf{k}}+\left(\partial/\partial\mathsf{q}_{\mathbf{k}}\right)\right],\\
\hat{A}_{\mathbf{k}}^{\dagger}= &\ v_{k}^{\ast}(\tilde{\tau})\left[-\tilde{\omega}_{\tilde{\tau},k}\hat{\mathsf{q}}_{\mathbf{k}}+\left(\partial/\partial\mathsf{q}_{\mathbf{k}}\right)\right].\label{annihilation2}
\end{align}
Using above operators we can define a number operator as 
\begin{equation}
\hat{N}_{k}=\hat{A}_{\mathbf{k}}^{\dagger}\hat{A}_{\mathbf{k}}=\left|v_{k}\right|^{2}\left[\left(\partial^{2}/\partial\mathsf{q}_{\mathbf{k}}^{2}\right)-\tilde{\omega}_{\tilde{\tau},k}^{2}\hat{\mathsf{q}}_{\mathbf{k}}^{2}\right],\label{annihilation3}
\end{equation}
such that 
\begin{equation}
\hat{N}_{k}\chi_{n}\left(\mathsf{q}_{\mathbf{k}},\tilde{\tau}\right)=n\chi_{n}\left(\mathsf{q}_{\mathbf{k}},\tilde{\tau}\right).
\end{equation}
The set $\{\chi_{n}\}$, given by Eq. \eqref{number-state}, forms
a complete orthonormal set for all $\tilde{\tau}$, and thus, we can
expand $|\tilde{0}\rangle$ in terms of these functions (evaluated
at $\tilde{\tau}_{0}$): 
\begin{equation}
|\tilde{0}\rangle=\sum_{n}b_{n}\chi_{n}\left(\mathsf{q}_{\mathbf{k}},\tilde{\tau}_{0}\right).
\end{equation}
Then, the expectation value of the number operator $\hat{N}_{k}$
with respect to $|\tilde{0}\rangle$ on the dressed background becomes
\begin{equation}
{\cal N}_{k}=\langle\tilde{0}|\hat{N}_{k}|\tilde{0}\rangle=\frac{1}{2}\left(\tilde{\omega}_{0}|v_{k}|^{2}+\frac{|v_{k}^{\prime}|^{2}}{\tilde{\omega}_{0}}-1\right),\label{Particle-Number}
\end{equation}
where $\tilde{\omega}_{0}=\tilde{\omega}_{\tilde{\tau},k}\left(\tilde{\tau}_{0}\right)$.

Within the above analysis of particle production on a general (anisotropic)
dressed background inspired by LQG, the ``in'' region can be assumed
to be at some instant in the Planck era where the background spacetime
is quantized due to LQG effects. The ``in'' vacuum state is thus
governed by an effective dressed geometry which resembles the Bianchi
type I model with components $\tilde{N}_{\phi}$ and $\tilde{a}_{i}$'s,
being the solutions of Eqs.~\eqref{eq:main-sys-1}-\eqref{eq:main-sys-3}.
The frequency of the field modes, $\tilde{\omega}_{0}=\tilde{\omega}_{\tilde{\tau},k}(\tilde{\tau}_{0})$,
is then given by Eq. \eqref{omega-BO} at some instant $\tilde{\tau}_{0}$,
say, at the initial quantum bounce, $\tilde{\tau}_{0}=\tilde{\tau}_{{\rm B}}$.
On the other hand, we assume that the ``out'' region is given by
the later classical de Sitter or FLRW phase due to the inflationary
scenario so that the frequency of the field modes $\tilde{\omega}_{\tilde{\tau},k}$
reduces to that of the field propagating on such a classical background
\begin{equation}
\tilde{\omega}_{\tilde{\tau},k}^{2}\left(\tilde{\tau}\right)=\omega_{\tau,k}^{2}\left(\tau\right)=k^{2}a^{4}+m^{2}a^{6}.\label{eq:later-class-FLRW}
\end{equation}
If the adiabaticity condition \eqref{adiabatic-condtion} is valid
for all $\tilde{\tau}\geq\tilde{\tau}_{0}$, we can expand the production
number \eqref{Particle-Number} for the general solution \eqref{WKB1}
as 
\begin{equation}
{\cal N}_{k}\approx\frac{1}{4}\left(\frac{\tilde{\omega}_{0}}{\tilde{\omega}_{\tilde{\tau},k}}+\frac{\tilde{\omega}_{\tilde{\tau},k}}{\tilde{\omega}_{0}}-2\right)+{\cal O}\left(\tilde{\omega}_{\tilde{\tau},k}^{\prime}/\tilde{\omega}_{\tilde{\tau},k}^{2}\right).\label{Particle-Number-WKB}
\end{equation}
To compute this expectation value \eqref{Particle-Number-WKB}, we
need to introduce relevant frequencies $\tilde{\omega}_{\tilde{\tau},k}$
and $\tilde{\omega}_{0}$ to be substituted in this equation. For
$\tilde{\omega}_{\tilde{\tau},k}$, we use the dispersion relation
\eqref{eq:later-class-FLRW}. This represents the dispersion relation
of a later classical FLRW epoch. For the initial frequency $\tilde{\omega}_{0}$,
we use the dispersion relation \eqref{omega-BO} which for all the
cases we studied turns out to be of the form \eqref{eq:mod-disp-rel}.
Then, \eqref{Particle-Number-WKB} becomes
\begin{align}
{\cal N}_{k}\approx &\ \frac{1}{4}\left\langle \hat{H}_{o}^{-1}\right\rangle ^{-\frac{1}{2}}\left[\frac{\left\langle \hat{H}_{o}^{-\frac{1}{2}}\hat{a}^{4}(\tau_{0})\hat{H}_{o}^{-\frac{1}{2}}\right\rangle }{a^{4}}+\frac{\left\langle \hat{H}_{o}^{-\frac{1}{2}}\hat{a}^{6}(\tau_{0})\hat{H}_{o}^{-\frac{1}{2}}\right\rangle }{a^{6}}\frac{m^{2}a^{2}}{k^{2}}\right]^{\frac{1}{2}}\left(1+\frac{m^{2}a^{2}}{k^{2}}\right)^{-\frac{1}{2}}\nonumber \\
 & +\frac{1}{4}\left\langle \hat{H}_{o}^{-1}\right\rangle ^{\frac{1}{2}}\left[\frac{\left\langle \hat{H}_{o}^{-\frac{1}{2}}\hat{a}^{4}(\tau_{0})\hat{H}_{o}^{-\frac{1}{2}}\right\rangle }{a^{4}}+\frac{\left\langle \hat{H}_{o}^{-\frac{1}{2}}\hat{a}^{6}(\tau_{0})\hat{H}_{o}^{-\frac{1}{2}}\right\rangle }{a^{6}}\frac{m^{2}a^{2}}{k^{2}}\right]^{-\frac{1}{2}}\left(1+\frac{m^{2}a^{2}}{k^{2}}\right)^{\frac{1}{2}}\nonumber \\
 & -\frac{1}{2}+{\cal O}\left(\tilde{\omega}_{k}^{\prime}/\tilde{\omega}_{k}^{2}\right).\label{Particle-Number2}
\end{align}
Eq. \eqref{Particle-Number2} indicates that when one or both of the
conditions 
\begin{align}
\frac{\left\langle \hat{H}_{o}^{-\frac{1}{2}}\hat{a}^{4}(\tau_{0})\hat{H}_{o}^{-\frac{1}{2}}\right\rangle }{\left\langle \hat{H}_{o}^{-1}\right\rangle }\neq &\ a^{4}(\tau_{0})\,,\label{condition-1}\\
\frac{\left\langle \hat{H}_{o}^{-\frac{1}{2}}\hat{a}^{6}(\tau_{0})\hat{H}_{o}^{-\frac{1}{2}}\right\rangle }{\left\langle \hat{H}_{o}^{-1}\right\rangle }\neq &\ a^{6}(\tau_{0})\,,\label{condition-2}
\end{align}
hold, then there will be particle creation. The details of the particle
creation of course depend on the explicit form of these expectation
values. These conditions are held even in the mean field approximation
where $\langle\hat{H}_{o}^{-\frac{1}{2}}\hat{a}^{4}(\tilde{\tau}_{0})\hat{H}_{o}^{-\frac{1}{2}}\rangle\sim\langle\hat{H}_{o}^{-1}\rangle\langle\hat{a}^{4}(\tilde{\tau}_{0})\rangle$
and $\langle\hat{H}_{o}^{-\frac{1}{2}}\hat{a}^{6}(\tilde{\tau}_{0})\hat{H}_{o}^{-\frac{1}{2}}\rangle\sim\langle\hat{H}_{o}^{-1}\rangle\langle\hat{a}^{6}(\tilde{\tau}_{0})\rangle$,
since $\langle\hat{a}^{4}(\tilde{\tau}_{0})\rangle\neq a^{4}$ and
$\langle\hat{a}^{6}(\tilde{\tau}_{0})\rangle\neq a^{6}$. Therefore,
the particle creation number \eqref{Particle-Number2} is always nonzero.
In comparison, in classical FLRW background, using the mechanism due
to  \citet{Berger:1975ag} (in a given adiabatic regime), there
is no particle creation for a \emph{massless} scalar field.

In computation of the particle number in this section, we have considered
only one mode of the field so that we could show the presence of nonzero
production rate for particles. In a more general case, in order to
obtain the total number of particle creation, one needs to consider
infinitely many modes of the scalar field propagating on the same
background dressed spacetime. As far as the field modes probe the
same background, sum over all modes can be done. This may pose a problem
that, in the presence of such modes, the energy density of the created
particles or that of the scalar field may fail to be negligible compared
to the energy density of the background quantum geometry which is
bounded by $0.41\rho_{{\rm Pl}}$ at the initial bounce. If the energy-momentum
of the field does become comparable to that in the background, then
we would not be able to neglect the back-reaction and our analysis
in subsection \ref{backreaction}, through the approximation $\Psi=\Psi_{o}\otimes\psi$,
fails to be self-consistent. This issue requires a careful treatment
of renormalization of the energy-momentum tensors of the quantum (inhomogeneous)
scalar field and the created scalar particles appearing in the theory.
In the next section (see subsection \ref{Field2} complemented by
appendices \ref{Ad-Vac} and \ref{adiabatic-energy-momentum}), we
will employ a renormalization method through the ``adiabatic regularization''
of the energy-momentum tensor in order to find a self-consistent solution
for the field modes.

\section{Amplitude of quantum fluctuations\label{fluctuation}}

Our aim in this section is to study the possible probes of the anisotropy
of the emergent dressed geometry $\tilde{g}_{ab}$ on which the scalar
field $\varphi(\phi,\mathbf{x})$, 
\begin{eqnarray}
\varphi(\phi,\mathbf{x}) & = & \frac{1}{(2\pi)^{3/2}}\int d^{3}\mathbf{k}\,\varphi_{\mathbf{k}}(\phi)e^{i\mathbf{k}\cdot\mathbf{x}}.\label{total-field1}
\end{eqnarray}
with massive modes $\mathsf{q}_{\mathbf{k}}$, propagates. 
We assume that the resultant solutions we obtained in subsection \ref{solution-Bianchi} for each field mode, are applicable to all modes. In other words, if one mode feels a dressed geometry $\tilde{g}_{ab}$, then all modes probe the same geometry $\tilde{g}_{ab}$, so that an integration over all modes can be possible on that background.

In subsection \ref{Field1}, we present the field equations on the
dressed Bianchi background and will discuss the possible solutions
for the frequency of the modes representing the anisotropy of the
dressed geometry near the Planck era. Then, by using this frequency
in subsection \ref{Field2}, we will investigate the possible solutions
$u_{k}$ to the field modes and compute the fluctuation amplitude
of the (auxiliary) field, 
\begin{equation}
\left\langle \tilde{0}\left|\hat{\chi}(\mathbf{x},\tilde{\eta})\,\hat{\chi}(\mathbf{y},\tilde{\eta})\right|\tilde{0}\right\rangle =\frac{1}{(2\pi)^{3}}\int d^{3}\mathbf{k}\,e^{\mathbf{k}\cdot(\mathbf{x}-\mathbf{y})}u_{k}^{\ast}(\tilde{\eta})u_{k}(\tilde{\eta}).\label{eq:fluctuation}
\end{equation}
Using this correlation function, we will show that an observable effect
from anisotropy of the dressed background can be extracted.

\subsection{Field equation on the dressed anisotropic background\label{Field1}}

In this subsection, we will discuss the modifications needed to characterize
the typical quantum fluctuations of the scalar modes on the smooth
effective background $\tilde{g}_{ab}$ in the Planck regime. Let us
follow the standard procedure used in cosmology literature and consider
the field modes propagating on the dressed Bianchi type-I background
with the metric
\begin{equation}
\tilde{g}_{ab}dx^{a}dx^{b}=-\tilde{N}_{\phi}^{2}\left(\phi\right)d\phi^{2}+\sum_{i=1}^{3}\tilde{a}_{i}^{2}\left(\phi\right)\left(dx^{i}\right)^{2},\label{BI-metric-class2}
\end{equation}
whose components are given by Eqs.~\eqref{eq:main-sys-1}-\eqref{eq:main-sys-3}
as we discussed in subsection \ref{solution-Bianchi}. By introducing
an auxiliary field $\chi_{\mathbf{k}}\coloneqq\tilde{c}^{1/2}(\tilde{\eta})\varphi_{\mathbf{k}}$
for a given mode $\mathbf{k}$, the corresponding Klein--Gordon equation
for that mode reads\footnote{see appendix \ref{field-equation-detail} for detailed description.}
\begin{equation}
\chi_{\mathbf{k}}^{\prime\prime}+\left[\tilde{\omega}_{\tilde{\eta},k}^{2}(\tilde{\eta})-\frac{\tilde{c}^{\prime\prime}}{2\tilde{c}}+\frac{\tilde{c}^{\prime2}}{4\tilde{c}^{2}}\right]\chi_{\mathbf{k}}=0,\label{field-eq-mode1-2a}
\end{equation}
where, we have defined the frequency $\tilde{\omega}_{\tilde{\eta},k}$
as 
\begin{equation}
\tilde{\omega}_{\tilde{\eta},k}^{2}(\tilde{\eta})=\tilde{c}\left(\sum_{i}^{3}\tilde{k}_{i}^{2}/\tilde{c}_{i}+\tilde{m}^{2}\right),\label{omega-BI-1}
\end{equation}
and 
\begin{equation}
\tilde{c}\coloneqq(\tilde{a}_{1}\tilde{a}_{2}\tilde{a}_{3})^{\frac{2}{3}}=(\tilde{c}_{1}\tilde{c}_{2}\tilde{c}_{3})^{\frac{1}{3}}\quad\quad{\rm and}\quad\quad\tilde{c}_{i}\coloneqq\tilde{a}_{i}^{2}.\label{c-tilde-2}
\end{equation}
In these equations, a prime denotes differentiation with respect to
the conformal time\footnote{At the fundamental level, evolution on quantum geometry is described
by the \emph{relational} time variable $\phi$ rather than cosmic
or conformal time. However, one can descend to a description in terms
of conformal time $\tilde{\eta}$ since $\tilde{g}_{ab}$, that incorporates
the quantum gravity corrections, is a smooth tensor field now.} $\tilde{\eta}$, which is defined by using the conformal lapse $N_{\tilde{\eta}}=(\tilde{a}_{1}\tilde{a}_{2}\tilde{a}_{3})^{1/3}=\tilde{c}^{1/2}$
in the relation $\tilde{N}_{\phi}d\phi=\tilde{N}_{\tilde{\eta}}d\tilde{\eta}$:
\begin{equation}
d\tilde{\eta}=\ell^{3}\langle\hat{H}_{o}^{-1}\rangle\tilde{c}(\phi)d\phi.\label{conformaltime}
\end{equation}
Note that the frequency (\ref{omega-BI-1}) we obtained in a conformal-time
gauge is different from the one we obtained earlier in Eq.~(\ref{omega-BO})
in a harmonic-time gauge by a factor $\tilde{c}^{-2}$. More precisely,
by using Eqs. \eqref{eq:main-sys-1} and \eqref{eq:main-sys-2}, the
frequency (\ref{omega-BI-1}) can be identically written as\footnote{We note that, as in the case of harmonic time gauge, since the frequency
(\ref{omega-BI-1}) can be written in terms of the \emph{original}
wave vector $k$, as given in Eq. (\ref{omega-BI-1-b}), we have denoted
the frequency by $\tilde{\omega}_{\tilde{\eta},k}$ instead of $\tilde{\omega}_{\tilde{\eta},\tilde{k}}$.} 
\begin{equation}
\tilde{\omega}_{\tilde{\eta},k}^{2}(\tilde{\eta})=\tilde{c}^{-2}\left[k^{2}\frac{\left\langle \hat{H}_{o}^{-\frac{1}{2}}\hat{a}^{4}\hat{H}_{o}^{-\frac{1}{2}}\right\rangle }{\langle\hat{H}_{o}^{-1}\rangle}+m^{2}\frac{\left\langle \hat{H}_{o}^{-\frac{1}{2}}\hat{a}^{6}\hat{H}_{o}^{-\frac{1}{2}}\right\rangle }{\langle\hat{H}_{o}^{-1}\rangle}\right]=\tilde{c}^{-2}\tilde{\omega}_{\tilde{\tau},k}^{2}.\label{omega-BI-1-b}
\end{equation}
This equation indicates that the possible alteration of the frequency
$\tilde{\omega}_{\tilde{\eta},k}$ can be achieved when various solutions
for $\tilde{c}$ are known. It is easy to show that there exist only
two classes of solutions for $\tilde{c}$, depending on whether or
not the mass of the field is dressed: 
\begin{enumerate}
\item In the case of the undressed mass (i.e., $\tilde{m}=m\neq0$), the
solution for $\tilde{c}$ is obtained from Eq.~(\ref{eq:main-sys-3})
simply as 
\begin{equation}
\tilde{c}(\tilde{\eta})=\frac{\left\langle \hat{H}_{o}^{-\frac{1}{2}}\hat{a}^{6}\hat{H}_{o}^{-\frac{1}{2}}\right\rangle ^{\frac{1}{3}}}{\langle\hat{H}_{o}^{-1}\rangle^{\frac{1}{3}}}.\label{ctilde-1}
\end{equation}
Hence, the frequency of the modes reads 
\begin{equation}
\tilde{\omega}_{\tilde{\eta},k}^{2}(\tilde{\eta})=k^{2}\frac{\left\langle \hat{H}_{o}^{-\frac{1}{2}}\hat{a}^{4}\hat{H}_{o}^{-\frac{1}{2}}\right\rangle }{\left\langle \hat{H}_{o}^{-\frac{1}{2}}\hat{a}^{6}\hat{H}_{o}^{-\frac{1}{2}}\right\rangle ^{\frac{2}{3}}\langle\hat{H}_{o}^{-1}\rangle^{\frac{1}{3}}}+m^{2}\frac{\left\langle \hat{H}_{o}^{-\frac{1}{2}}\hat{a}^{6}\hat{H}_{o}^{-\frac{1}{2}}\right\rangle ^{\frac{1}{3}}}{\langle\hat{H}_{o}^{-1}\rangle^{\frac{1}{3}}}.\label{omega-BI-1b}
\end{equation}
\item For the modes with the dressed mass or the case of massless modes,
using  Eq. (\ref{anisotropy-scale1}) or Eqs. (\ref{a1a2k3-1})
and (\ref{a1a2k3-3}), we get 
\begin{equation}
\tilde{c}(\tilde{\eta})=\xi^{2}\frac{\left\langle \hat{H}_{o}^{-\frac{1}{2}}\hat{a}^{4}\hat{H}_{o}^{-\frac{1}{2}}\right\rangle ^{\frac{1}{2}}}{\langle\hat{H}_{o}^{-1}\rangle^{\frac{1}{2}}}=\xi^{2}\tilde{a}^{2}(\tilde{\eta}),\label{ctilde-2}
\end{equation}
where $\xi$ is a parameter that distinguishes two different solutions
for $\tilde{c}$; for $\xi^{2}=1$ the solution of $\tilde{c}$ associates
the one provided by the case 1 (Eq.~(\ref{anisotropy-scale1b})),
and the value $\xi^{2}\equiv(\rho_{12}\rho_{23}^{2}\rho_{33}^{3})^{-1/3}$
denotes the solution of the case 2 (Eqs.~(\ref{a1a2k3-1}) and (\ref{a1a2k3-3})).
Moreover, $\tilde{a}$ was defined in  Eq.~(\ref{anisotropy-scale1b})
as well as Eq.~(\ref{massless-dressed-scale-1}). For this relation
of $\tilde{c}$, the frequency becomes 
\begin{equation}
\tilde{\omega}_{\tilde{\eta},k}^{2}(\tilde{\eta})=k^{2}+\tilde{m}^{2}\xi^{2}\tilde{a}^{2}.\label{omega-BI-1c}
\end{equation}
This is equal to the frequency of the modes with a dressed mass $\tilde{m}$
given by (\ref{dressed-mass0}) propagating on the dressed isotropic
background with the scale factor $\xi\tilde{a}(\tilde{\eta})$. This
class of the solutions, as introduced in subsection \ref{dressedmass-sol},
are identical to the isotropic solutions with $\xi=1$, given earlier
in Refs.~\cite{Ashtekar:2009mb,Agullo:2012fc,Agullo:2013ai} for
massive and massless modes. 
\end{enumerate}
Our analyses in this subsection implies that despite the dependence
of the anisotropic solutions we have found in subsection \ref{solution-Bianchi}
for the effective dressed background, the emergent frequencies of
the modes are independent of the anisotropic parameters encoded in
$\alpha_{ij},\beta_{ij},\gamma_{ij},\sigma_{ij}$. Nevertheless, the
frequency (\ref{omega-BI-1b}) of the field modes propagating on the
anisotropic dressed spacetime is different from those propagating
on the isotropic dressed geometry (i.e., those associated to the frequency
(\ref{omega-BI-1c})). Therefore, the frequency (\ref{omega-BI-1b})
distinguishes the amplitude of the quantum fluctuations for the fields
propagating on the smooth quantum gravity induced anisotropic background
from those propagating on the isotropic spacetime, and can provide
a probing for the quantum gravitational nature of the primordial anisotropy
from the deep Planck regime. In the next subsection, we will study
the amplitude of the quantum fluctuations of the field on the anisotropic
dressed background which can provide an observable signature for the
quantum gravity inspired anisotropy in the early Universe.

\subsection{The adiabatic condition on vacuum states\label{Field2}}

In this subsection we compute the power spectrum of the quantum fluctuations,
using general mode solutions $u_{k}(\tilde{\eta})$ of the Klein--Gordon
equation (\ref{field-eq-mode1-2a}). These modes\footnote{Because of the specific symmetry emerged herein our model, due to
the frequencies (\ref{omega-BI-1b}) and (\ref{omega-BI-1c}), the
subscription of the mode function $u_{\tilde{\mathbf{k}}}$ can be
written as $u_{\mathbf{k}}$ or $u_{k}$. With a similar argument,
we can write $a_{\mathbf{k}}$ instead of $a_{\tilde{\mathbf{k}}}$.
Therefore, we have written the field equation and their possible solutions
in terms of $k$, $\tilde{c}$ and $\tilde{m}$ and the isotropic
quantum geometry fluctuations. Within this expression, the integration
(e.g., given by Eq.~\ref{total-field1} or Eq.~(\ref{eq:fluctuation}))
over all modes $k$, instead of $\tilde{k}$ is relevant.} $u_{k}$ obey the equation $u_{k}^{\prime\prime}+(\tilde{\omega}_{\tilde{\eta},k}^{2}-{\cal Q})u_{k}=0$,
and form a complete set of orthonormal basis with their complex conjugates
$u_{k}^{\ast}$ under the scalar product 
\begin{equation}
W(u_{k}^{\ast},u_{k})\coloneqq u_{k}u_{k}^{\ast\prime}-u_{k}^{\ast}u_{k}^{\prime}=2i.\label{Wronskian}
\end{equation}
Thus, any solution $\chi_{\mathbf{k}}(\tilde{\eta})$ of Eq. (\ref{field-eq-mode1-2a})
with the same $\mathbf{k}$ can be expanded as 
\begin{equation}
\chi_{\mathbf{k}}(\tilde{\eta})=\frac{1}{\sqrt{2}}\left[a_{\mathbf{k}}u_{k}^{\ast}(\tilde{\eta})+a_{-\mathbf{k}}^{\ast}u_{k}(\tilde{\eta})\right],\label{mode-decompose1}
\end{equation}
where $a_{\mathbf{k}}$ and $a_{\mathbf{k}}^{\ast}$ are constants
of integration\footnote{The normalization condition (\ref{Wronskian}) for $\chi_{\mathbf{k}}$
provides that the corresponding annihilation and creation operators
$\hat{a}_{\mathbf{k}}$ and $\hat{a}_{-\mathbf{k}}^{\dagger}$ satisfy
the canonical commutation relations in quantum theory.}, which in quantum theory become annihilation and creation operators
for each mode $\chi_{\mathbf{k}}$. With normalization (\ref{Wronskian}),
operators $\hat{a}_{\mathbf{k}}$ and $\hat{a}_{\mathbf{k}}^{\dagger}$
will satisfy the commutation relation 
\begin{equation}
\left[\hat{a}_{\mathbf{k}},\hat{a}_{\mathbf{k}^{\prime}}^{\dagger}\right]=\hbar\ell^{3}\delta_{\mathbf{k},\mathbf{k}^{\prime}},
\end{equation}
with all other commutation relations vanishing. A choice of basis
with \emph{positive frequency} solutions $u_{k}(\tilde{\eta})$ determines
a vacuum state $|0\rangle$, which can be defined as the eigenstate
of the annihilation operators with vanishing eigenvalue, i.e., $\hat{a}_{\mathbf{k}}|0\rangle=0$.
Consequently, a Fock space is generated for the quantum theory by
repeatedly acting on the vacuum by creation operators $\hat{a}_{\mathbf{k}}^{\dagger}$. 

Using the reality condition for the scalar field, $\varphi_{\mathbf{k}}=\varphi_{-\mathbf{k}}^{\ast}$,
and since $\varphi_{\mathbf{k}}=\tilde{c}^{-1/2}\chi_{\mathbf{k}}$,
we obtain $\chi_{\mathbf{k}}^{*}(\tilde{\eta})=\chi_{-\mathbf{k}}(\tilde{\eta})$.
Hence, the general solution for the (auxiliary) scalar field satisfying
the Klein--Gordon equation can be expressed by 
\begin{equation}
\chi(\tilde{\eta},\mathbf{x})=\frac{1}{\ell^{3}}\sum_{\mathbf{k}\in\mathcal{L}}\chi_{\mathbf{k}}(\tilde{\eta},\mathbf{x}),\label{mode-decompose2}
\end{equation}
as the sum of all modes of the field $\chi_{\mathbf{k}}(\tilde{\eta},\mathbf{x})$
given by 
\begin{equation}
\chi_{\mathbf{k}}=\frac{1}{\sqrt{2}}\left[a_{\mathbf{k}}u_{k}^{*}(\tilde{\eta})e^{i\mathbf{k}\cdot\mathbf{x}}+a_{\mathbf{k}}^{\ast}u_{k}(\tilde{\eta})e^{-i\mathbf{k}\cdot\mathbf{x}}\right].\label{mode-decompose2b}
\end{equation}
Note that in the above equation, we have changed the integration variable
$\mathbf{k}\rightarrow-\mathbf{k}$.

In quantum theory of scalar field on the dressed geometry (\ref{BI-metric-class2})
in the Planck era with $\tilde{\eta}\lesssim\eta_{p}$, we can construct
a Fock space ${\cal H}_{{\rm F}}$ by defining a vacuum state $|0\rangle$,
which is associated to the positive frequency solution $u_{k}(\tilde{\eta})$
of Eq.~(\ref{field-eq-mode1-2a}): 
\begin{equation}
u_{k}^{\prime\prime}+\Big(\tilde{\omega}_{\tilde{\eta},k}^{2}(\tilde{\eta})-{\cal Q}(\tilde{\eta})\Big)u_{k}=0,\label{KG-1-eff-2}
\end{equation}
where, ${\cal Q}\equiv\tilde{c}^{\prime\prime}/2\tilde{c}-\tilde{c}^{\prime2}/4\tilde{c}^{2}$
and $u_{k}$ satisfies the normalization condition (\ref{Wronskian}).
Different families of solutions provide different definitions of the
vacuum state. Under these conditions, we can determine evolution of
the quantum fields on the quantum gravity induced dressed spacetime.
The parameter ${\cal Q}(\tilde{\eta})$ has a linear dependence on
the curvature of $\tilde{g}_{ab}$ and introduces a physical length
$L(\tilde{\eta})$. For the frequency $\Omega_{k}^{2}(\tilde{\eta})\equiv\tilde{\omega}_{\tilde{\eta},k}^{2}(\tilde{\eta})-{\cal Q}(\tilde{\eta})$
to be positive, the wave-number must always satisfy an inequality
$k^{2}\geq k_{\star}^{2}$, where 
\begin{equation}
k_{\star}^{2}(\tilde{\eta})\coloneqq{\cal Q}(\tilde{\eta})\tilde{c}^{2}(\tilde{\eta})\langle\hat{H}_{o}^{-1}\rangle-m^{2}\frac{\left\langle \hat{H}_{o}^{-\frac{1}{2}}\hat{a}^{6}\hat{H}_{o}^{-\frac{1}{2}}\right\rangle }{\left\langle \hat{H}_{o}^{-\frac{1}{2}}\hat{a}^{4}\hat{H}_{o}^{-\frac{1}{2}}\right\rangle }.
\end{equation}
In this case the resulting vacuum state will be well-defined for modes
$u_{k}$ with wavelengths shorter than the length scale $L(\tilde{\eta})=\sqrt{\tilde{c}(\tilde{\eta})}/k_{\star}$.
Hence, modes with large momenta, $k/\sqrt{\tilde{c}}\gg1/L$, describe
the vacuum in short distances. This respects the regularity condition
for a natural choice of the mode function in the \emph{ultra-violet}
regime with $(\sqrt{\tilde{c}}/kL)\rightarrow0$, where the curvature
has negligible effects and solutions to (\ref{KG-1-eff-2}) reduce
to the standard mode functions $e^{-ik\eta}/\sqrt{2k}$ in Minkowski
space. This regime constitutes a limit of arbitrary slow time variation
of the metric functions $\tilde{a}_{i}(\tilde{\eta})$ with respect
to the time $\tilde{\eta}$, which is the so-called \emph{adiabatic}
regime.

Our aim is to determine the mode functions that describe the physical
vacuum and particles. To define the adiabatic vacuum modes, one can
employ a positive-frequency generalized WKB method \cite{Parker:1974qw}
to get 
\begin{equation}
\underline{u}_{k}(\tilde{\eta})=\frac{1}{\sqrt{W_{k}(\tilde{\eta})}}\exp\Big(-i\int^{\tilde{\eta}}W_{k}(\eta)d\eta\Big),\label{adibatic-1}
\end{equation}
which yields an approximate solution to the Klein--Gordon equation
(\ref{KG-1-eff-2}). Note that $\underline{u}_{k}(\tilde{\eta})$
are guaranteed to satisfy the Wronskian condition (\ref{Wronskian})
for any real, non-negative function $W_{k}(\tilde{\eta})$. An appropriate
function $W_{k}(\tilde{\eta})$ is given by the method of Chakraborty \cite{Chakraborty:1973}
as 
\begin{equation}
W_{k}(\tilde{\eta})=\big[Y(1+\underline{\epsilon_{2}})(1+\underline{\epsilon_{4}})\big]^{\frac{1}{2}},\label{W-k-1}
\end{equation}
in which we have set $Y\equiv\Omega_{k}^{2}=\tilde{\omega}_{\tilde{\eta},k}^{2}-{\cal Q}$,
and 
\begin{align}
\underline{\epsilon_{2}} & \coloneqq-Y^{-\frac{3}{4}}\partial_{\tilde{\eta}}\Big(Y^{-\frac{1}{2}}\partial_{\tilde{\eta}}Y^{\frac{1}{4}}\Big),\label{epsilon-under-1}\\
\underline{\epsilon_{4}} & \coloneqq-Y^{-\frac{1}{2}}(1+\underline{\epsilon_{2}})^{-\frac{3}{4}}\partial_{\tilde{\eta}}\Big\{[Y(1+\underline{\epsilon_{2}})]^{-\frac{1}{2}}\partial_{\tilde{\eta}}(1+\underline{\epsilon_{2}})^{\frac{1}{4}}\Big\}.\label{epsilon-under-2}
\end{align}
If $\underline{u}_{k}(\tilde{\eta})$ is a solution for the exact
mode function satisfying the Klein--Gordon equation (\ref{KG-1-eff-2}),
then $W(\tilde{\eta})$ is required to satisfy the relation (\ref{W-k-1}).
However, instead of solving for $W_{k}$ exactly, one can generate
asymptotic series in orders of time derivatives (with respect to $\tilde{\eta}$)
of the background metric. Terminating this series at a given order
will define an adiabatic mode $\underline{u}_{k}(\tilde{\eta})$ to
such order. Thus, up to an order $n$, denoted by ${\cal O}(\sqrt{\tilde{c}}/kL_{n+\varepsilon})^{n+\varepsilon}$
(being power of time variation of the $\tilde{a}_{i}$ components
of the metric), the asymptotic adiabatic expansion of the frequency
$W_{k}$ is defined to match $W_{k}(\tilde{\eta})=W_{k}^{(0)}+W_{k}^{(2)}+\cdot\cdot\cdot+W_{k}^{(n)}$,
which is obtained for higher order estimates by iteration. Since the
mode functions are specified by the adiabatic expansion scheme (\ref{adibatic-1}),
the expectation values of the energy-momentum of the field, $\langle\tilde{0}|\hat{T}_{ab}|\tilde{0}\rangle$
with respect to the adiabatic vacuum $|\tilde{0}\rangle$ can be computed.
It can be shown that all ultra-violet divergences are contained in
terms of adiabatic order equal to and smaller than four \cite{Fulling:1974pu}
(see appendix \ref{adiabatic-energy-momentum}). Therefore, we restrict
ourselves to the fourth order adiabatic states by setting $n=4$,
and define $W_{k}(\tilde{\eta})$ to match the terms in 
\begin{equation}
W_{k}(\tilde{\eta})=W_{k}^{(0)}+W_{k}^{(2)}+W_{k}^{(4)},\label{initial-con3-a}
\end{equation}
that fall slowly in $\tilde{\omega}_{\tilde{\eta},k}$, rather than
demanding that $\underline{u}_{k}(\tilde{\eta})$ satisfies the exact
mode equation (\ref{KG-1-eff-2}). At the appropriate rate (say, at
the initial quantum bounce\footnote{The problem of initial conditions in the early universe is still under
discussion, which might be solved in the framework of a quantum theory
of gravity. Therefore, a natural choice for the preferred instant
of time in LQC is provided by the quantum bounce, at $\tilde{\eta}=\tilde{\eta}_{b}$.} $\tilde{\eta}=\tilde{\eta}_{b}$), given by the asymptotic conditions
\begin{align}
|u_{k}(\tilde{\eta}_{b})|= &\ \big|\underline{u}_{k}(\tilde{\eta}_{b})\big|\Big(1+{\cal O}\big(\sqrt{\tilde{c}}/kL_{4+\varepsilon}\big)^{4+\varepsilon}\Big),\nonumber \\
|u_{k}^{\prime}(\tilde{\eta}_{b})|= &\ \big|\underline{u}_{k}^{\prime}(\tilde{\eta}_{b})\big|\Big(1+{\cal O}\big(\sqrt{\tilde{c}}/kL_{4+\varepsilon}\big)^{4+\varepsilon}\Big),\label{adiab-cond2}
\end{align}
(with positive real number $\varepsilon$), the field's exact mode
functions $u_{k}(\tilde{\eta})$ match the adiabatic functions $\underline{u}_{k}(\tilde{\eta})$
(up to the order four). If the conditions (\ref{adiab-cond2}) are
held for the mode functions at some initial time $\tilde{\eta}_{b}$,
they will be held for all times $\tilde{\eta}$. That is, an observable
vacuum state associated to $u_{k}(\tilde{\eta})$, given at any time
$\tilde{\eta}$, will be of the 4th order.

By discarding terms of adiabatic order higher than four in (\ref{W-k-1})
we obtain 
\begin{equation}
W_{k}(\tilde{\eta})=\tilde{\omega}_{\tilde{\eta},k}(1+\epsilon_{2}+\epsilon_{4})^{\frac{1}{2}},\label{adibatic-2}
\end{equation}
where the $k$-dependent parameters $\epsilon_{2}$, $\epsilon_{4}$
are defined as 
\begin{align}
\epsilon_{2} & =\underline{\epsilon_{2}}-\tilde{\omega}_{\tilde{\eta},k}^{-2}{\cal Q},\label{adibatic-2-sec1}\\
\epsilon_{4} & =\underline{\epsilon_{4}}-\underline{\epsilon_{2}}\tilde{\omega}_{\tilde{\eta},k}^{-2}{\cal Q}.\label{adibatic-2-sec}
\end{align}
The leading order term of $\underline{\epsilon_{2}}$ is of second
order, whereas the leading order in $\underline{\epsilon_{4}}$ is
four. Thus, $\epsilon_{2}$ contains terms of orders two, four and
higher, while $\epsilon_{4}$ contains leading order term four. Of
course we will consider only terms until the fourth order in their
expressions. Notice that we have dropped terms like $\underline{\epsilon_{2}}\underline{\epsilon_{4}}$
because they are of sixth or higher order. For detailed expressions
of $\epsilon_{2}$ and $\epsilon_{4}$ see appendix \ref{Ad-Vac}.

Once the vacuum state $|\tilde{0}\rangle$ of the fourth adiabatic
order is determined by a set of mode functions $u_{k}(\tilde{\eta})$,
following the mechanism we discussed above, we can write the power
spectrum of quantum fluctuations using the equal-time correlation
function (\ref{eq:fluctuation}) as 
\[
{\cal P}_{\chi}(k)=\frac{2\pi^{2}}{k^{3}}\left\langle \tilde{0}\left|\hat{\chi}(\mathbf{x},\tilde{\eta})\,\hat{\chi}(\mathbf{y},\tilde{\eta})\right|\tilde{0}\right\rangle =|u_{k}(\tilde{\eta})|^{2}.
\]
This implies that, in order to calculate the ${\cal P}_{\chi}(k)$,
one needs to evolve the set of mode function $u_{k}$. As we have
discussed earlier, we are concerned about the anisotropy generated
by quantum fluctuation of the scalar field in the emergent dressed
background from the Planck era. Therefore, once the mode functions
$u_{k}(\tilde{\eta})$ associated to the vacuum states are determined
for the isotropic and the anisotropic emergent geometries, by a comparison
between the power spectrum of the quantum fluctuations due to our
analysis and the data provided by observations, one can clarify the
nature of the CMB anisotropies (see e.g. \cite{Komatsu:2010fb})
which may arise from the quantum gravity effects from the Planck era.

In particular, from Eq.~(\ref{v-vprime1}), we get the expression
for the mode function $|u_{k}|$ up to fourth adiabatic order. In
terms of this mode solution, the power spectrum of the quantum fluctuations
reads 
\begin{equation}
{\cal P}_{\chi}(k)=\tilde{\omega}_{\tilde{\eta},k}^{-1}\Big[1-\frac{1}{2}\big(\epsilon_{2}^{(2)}+\epsilon_{2}^{(4)}+\epsilon_{4}^{(4)}\big)+\frac{3}{8}\big(\epsilon_{2}^{(2)}\big)^{2}\Big].\label{Power-Spectrum}
\end{equation}
This implies that the power spectra ${\cal P}_{\chi}(k)$ arising
on the dressed Bianchi type I background (\ref{BI-metric-class2})
depend on the anisotropy encoded in the leading adiabatic terms including
$\tilde{\omega}_{\tilde{\eta},k}(\tilde{\eta})$ and ${\cal Q}(\tilde{\eta})$
and their (conformal) time derivatives.

In order to make a comparison between the anisotropic power spectrum
with that of the isotropic one, dressed background, let us rewrite
the general frequency (\ref{omega-BI-1-b}) of the modes on the emergent,
dressed Bianchi as 
\begin{equation}
\tilde{\omega}_{\tilde{\eta},k}^{2}(\tilde{\eta})=\frac{\tilde{a}^{4}}{\tilde{c}^{2}}\left[k^{2}+\tilde{m}^{2}\tilde{a}^{2}\right].
\end{equation}
The anisotropy factor $\tilde{c}$ is given either by Eq.~(\ref{ctilde-1})
or Eq.~(\ref{ctilde-2}) depending on the anisotropic solution we
have chosen. Let us now denote for convenience, the term in the bracket
by $\tilde{\omega}_{I}^{2}$ as the frequency associated to the mode
$k$ (with the dressed mass $\tilde{m}$) propagating on the isotropic
background with the scale factor $\tilde{a}$. Then, denoting the
left hand side by $\tilde{\omega}_{A}^{2}$ as the frequency of the
same massive mode on the anisotropic background (\ref{BI-metric-class2}),
we write the relation between two frequencies as 
\begin{equation}
\tilde{\omega}_{A}^{2}(\tilde{\eta})=A^{2}(\tilde{\eta})\tilde{\omega}_{I}^{2}(\tilde{\eta}),\quad{\rm where}\quad A(\tilde{\eta}):=\tilde{a}^{2}(\tilde{\eta})\tilde{c}^{-1}(\tilde{\eta}).
\end{equation}
In the isotropic limit, we have $A=1$. The expression for the power
spectra (\ref{Power-Spectrum}) constitutes of the frequency $\tilde{\omega}_{A}$
and its (conformal) time derivatives. That is, it contains power terms
of $\tilde{\omega}_{A}$ and its derivatives up to fourth order. Therefore,
by assuming that the changes in the quantum fluctuations of the geometry
associated to the anisotropy parameter $A(\tilde{\eta})$ are slow,
we can neglect terms like $A^{\prime}/A$ and its higher time derivatives.
Using this, the power spectra (\ref{Power-Spectrum}) reduce to 
\begin{eqnarray}
{\cal P}_{\chi}^{(A)}(k)\  & \approx & \ \tilde{\omega}_{A}^{-1}\left[1+\frac{1}{2}\tilde{\omega}_{A}^{-2}{\cal Q}_{A}-\frac{1}{8}\tilde{\omega}_{A}^{-4}{\cal Q}_{A}^{\prime\prime}+\frac{3}{8}\tilde{\omega}_{A}^{-4}{\cal Q}_{A}^{2}\right]\nonumber \\
 &  & +~{\rm The~terms~including}~(\partial A/A),
\end{eqnarray}
where, $\partial A$ holds for the derivatives of $A(\tilde{\eta})$
with respect to $\tilde{\eta}$ including first order derivative,
second order derivative and so on. In terms of the power spectrum
of the modes on the isotropic case, we can write 
\begin{eqnarray}
\frac{{\cal P}_{\chi}^{(A)}}{{\cal P}_{\chi}^{(I)}}=A^{-1}(\tilde{\eta})\left[\frac{1+\frac{1}{2}A^{-2}\tilde{\omega}_{I}^{-2}{\cal Q}_{A}-\frac{1}{8}A^{-4}\tilde{\omega}_{I}^{-4}{\cal Q}_{A}^{\prime\prime}+\frac{3}{8}A^{-2}\tilde{\omega}_{I}^{-4}{\cal Q}_{A}^{2}}{1+\frac{1}{2}\tilde{\omega}_{I}^{-2}{\cal Q}_{I}-\frac{1}{8}\tilde{\omega}_{I}^{-4}{\cal Q}_{I}^{\prime\prime}+\frac{3}{8}\tilde{\omega}_{I}^{-4}{\cal Q}_{I}^{2}}\right],
\end{eqnarray}
where, ${\cal Q}_{A}=\tilde{c}^{\prime\prime}/2\tilde{c}-\tilde{c}^{\prime2}/4\tilde{c}^{2}$
and ${\cal Q}_{I}=\tilde{a}^{\prime\prime}/\tilde{a}$.

It should be noted that, although the power spectrum we obtained for
the anisotropic background would be different from the one for the
isotropic geometry, the power spectrum we obtained depends only on
the product $\tilde{a}_{1}\tilde{a}_{2}\tilde{a}_{3}$ and not on
any specific scale factor $\tilde{a}_{i}$. The above expression can
in principle be used to predict observational signatures imprinted
on the power spectrum of the cosmic microwave background, which can
distinguish between the isotropic and anisotropic cases.

\subsection{The energy density of the created particles\label{Particle-Planck}}

In order to complete the discussion of the energy density of particle
creations we presented in section \ref{pairprod} when considering
infinitely many modes of the field, we provide an analysis of particle
energy regularization process in this subsection.

The Fock space ${\cal H}_{{\rm F}}$ for the quantum field $\varphi$
propagating on the dressed Bianchi I background (\ref{BI-metric-class2})
is constructed from a vacuum state $|\tilde{0}\rangle$, which is
determined by a choice of the positive frequency solution $v_{k}(\tilde{\eta})$
satisfying Eq.~(\ref{KG-1-eff-2}). Then, any two solutions $v_{k}$
and $\underline{v}_{k}$ of (\ref{KG-1-eff-2}) are related through
the time-independent Bogolyubov coefficients $\alpha_{k}$ and $\beta_{k}$
by 
\begin{equation}
\underline{v}_{k}(\tilde{\eta})=\alpha_{k}v_{k}(\tilde{\eta})+\beta_{k}v_{k}^{\ast}(\tilde{\eta}).\label{Bog-relation1}
\end{equation}
Since $v_{k}$ and $\underline{v}_{k}$ are normalized due to the
condition (\ref{Wronskian}), the coefficients $\alpha_{k}$ and $\beta_{k}$
satisfy the relation $|\alpha_{k}|^{2}-|\beta_{k}|^{2}=1$. By substituting
both sides of Eq.~(\ref{Bog-relation1}) into (\ref{mode-decompose2b})
one obtains a relation between creation and annihilation operators
associated with two families of mode function, as 
\begin{equation}
\hat{a}_{\mathbf{k}}=\alpha_{k}\underline{\hat{a}}_{\mathbf{k}}+\beta_{k}^{\ast}\underline{\hat{a}}_{\mathbf{k}}^{\dagger}.
\end{equation}
In the Heisenberg picture, the initial vacuum state $|\tilde{0}\rangle$
is the vacuum state of the system for all times. The physical number
operator $\hat{N}_{\mathbf{k}}$ (see Eq.~(\ref{annihilation3}))
which counts excitation/particles of $v_{k}$ modes, yields  the
average number of particles in the $|\tilde{0}\rangle$ vacuum, associated
to the ``under-barred'' modes,
\begin{equation}
{\cal N}_{k}\coloneqq\langle\tilde{0}|\hat{N}_{\mathbf{k}}|\tilde{0}\rangle=(\hbar\ell^{3})^{-1}\langle\tilde{0}|\hat{a}_{\mathbf{k}}^{\dagger}\hat{a}_{\mathbf{k}}|\tilde{0}\rangle=|\beta_{k}|^{2}.
\end{equation}
In other words, the vacuum state associated to the $v_{k}$-mode contains
${\cal N}_{k}=|\beta_{k}|^{2}$ particles in the vacuum state associated
to $\underline{v}_{k}$-mode.

We may consider the solution of Eq.~(\ref{KG-1-eff-2}) in the WKB
form \cite{Hu:1974fs,Zeldovich:1971mw} as 
\begin{equation}
u_{k}(\tilde{\eta})=\frac{1}{\sqrt{\tilde{\omega}_{k,\tilde{\eta}}}}\Big[\alpha_{k}e_{k}(\tilde{\eta})+\beta_{k}e_{k}^{\ast}(\tilde{\eta})\Big],\label{beta-particle-1}
\end{equation}
where 
\begin{equation}
e_{k}(\tilde{\eta})\coloneqq\exp\Big(-i\int^{\tilde{\eta}}\tilde{\omega}_{k,\tilde{\eta}}(\eta)d\eta\Big).
\end{equation}
With the introduction of functions $\alpha_{k}$ and $\beta_{k}$,
we have the freedom of imposing an additional condition on the time
($\tilde{\eta}$) derivative of $u_{k}$ such that 
\begin{equation}
u_{k}^{\prime}(\tilde{\eta})=-i\sqrt{\tilde{\omega}_{k,\tilde{\eta}}}\Big[\alpha_{k}e_{k}(\tilde{\eta})-\beta_{k}e_{k}^{\ast}(\tilde{\eta})\Big].\label{beta-particle-2}
\end{equation}
Then, from the Wronskian condition (\ref{Wronskian}) on $u_{k}$,
we obtain $|\alpha_{k}|^{2}-|\beta_{k}|^{2}=1$. Inverting  Eqs.~(\ref{beta-particle-1})
and (\ref{beta-particle-2}) yields 
\begin{equation}
\beta_{k}=\frac{\sqrt{\tilde{\omega}_{k,\tilde{\eta}}}}{2}\left(u_{k}-\frac{i}{\tilde{\omega}_{k,\tilde{\eta}}}u_{k}^{\prime}\right)e_{k}.\label{beta_rel1}
\end{equation}
We fix the initial condition $\alpha_{k}=1$ and $\beta_{k}=0$ at
the quantum bounce so that no particle has  been created at $\tilde{\eta}=\tilde{\eta}_{b}$.
Inserting this into Eqs.~(\ref{beta-particle-1}) and (\ref{beta-particle-2})
yields $u_{k}(\tilde{\eta}_{b})=1/\tilde{\omega}_{k,\tilde{\eta}}(\tilde{\eta}_{b})$,
where $e_{k}(\tilde{\eta}_{b})=1$, and $u_{k}^{\prime}(\tilde{\eta}_{b})=-i\tilde{\omega}_{k,\tilde{\eta}}(\tilde{\eta}_{b})u_{k}(\tilde{\eta}_{b})$.
By making use of relation (\ref{Wronskian}), we find for the particle
production 
\begin{equation}
{\cal N}_{k}=\frac{1}{4}\Big(\tilde{\omega}_{k,\tilde{\eta}}|u_{k}|^{2}+\tilde{\omega}_{k,\tilde{\eta}}^{-1}|u_{k}^{\prime}|^{2}-2\Big).\label{Part-number}
\end{equation}
As expected, this equation is identical to the equation (\ref{Particle-Number})
for the particle creation rate we obtained in harmonic-time gauge\footnote{Bear in mind that there is a difference in the factor of the square
root of the denominator of the WKB expansion (\ref{WKB1}) which made
the particle number (\ref{Particle-Number}) twice the one derived
in Eq.~(\ref{Part-number}). Moreover, the frequency in the harmonic
time gauge has a difference of factor $\tilde{c}^{2}$ with respect
to the frequency (\ref{omega-BI-1-b}) given in the conformal time
gauge which can be disappeared by setting the new mode function solution
$u(\tilde{\eta})$ rather than $v(\tilde{\tau})$ in Eq.~(\ref{Part-number}).}.

The energy density of created particles for infinitely many modes
is 
\begin{equation}
\rho_{{\rm par}}=\frac{1}{\ell^{3}}\sum_{\mathbf{k}}\varrho_{k}(\tilde{\eta}),\label{energy-part-tot}
\end{equation}
where $\varrho_{k}$ is the energy density of each mode 
\begin{align}
\varrho_{k}(\tilde{\eta})\coloneqq &\ \tilde{\omega}_{k,\tilde{\eta}}{\cal N}_{k}\nonumber \\
= &\ \frac{1}{4}\Big(|u_{k}^{\prime}|^{2}+\tilde{\omega}_{k,\tilde{\eta}}^{2}|u_{k}|^{2}-2\tilde{\omega}_{k,\tilde{\eta}}\Big).\label{energy-part-tot1}
\end{align}
Notice that at the initial quantum bounce $\tilde{\eta}=\tilde{\eta}_{b}$,
one obtains $\varrho_{k}(\tilde{\eta}_{b})=0$.

Our aim is to derive an expression for the number of particle ${\cal N}_{k}$
in the adiabatic regime. By substituting the mode functions $\underline{u}_{k}$
and their derivatives, $\underline{u}_{k}^{\prime}$, from Eqs.~(\ref{v-vprime1})
and (\ref{v-vprime2}) into expression above, we obtain the number
of particles up to the fourth adiabatic order as 
\[
{\cal N}_{k}={\cal N}_{k}^{(0)}+{\cal N}_{k}^{(2)}+{\cal N}_{k}^{(4)},
\]
where, the zeroth, second and fourth adiabatic orders are 
\begin{equation}
{\cal N}_{k}^{(0)}=0,\quad\quad{\cal N}_{k}^{(2)}=\frac{1}{16}\frac{(\tilde{\omega}_{k,\tilde{\eta}}^{\prime})^{2}}{\tilde{\omega}_{k,\tilde{\eta}}^{4}},\quad\quad{\cal N}_{k}^{(4)}=\frac{1}{16}\left[\big(\epsilon_{2}^{(2)}\big)^{2}+\frac{\tilde{\omega}_{k,\tilde{\eta}}^{\prime}}{\tilde{\omega}_{k,\tilde{\eta}}^{3}}\epsilon_{2}^{\prime(3)}-\frac{(\tilde{\omega}_{k,\tilde{\eta}}^{\prime})^{2}}{2\tilde{\omega}_{k,\tilde{\eta}}^{4}}\epsilon_{2}^{(2)}\right].\label{Number-adiabatic-1}
\end{equation}
This equation implies that no particle production occurs in the zeroth
order adiabatic expansion of the mode function (or superadiabatic
regime). Particles are created only for expansions of adiabatic series
equal to or more than two. In the very short scales, where $m\ll k$,
from second relation in (\ref{Number-adiabatic-1}) we get the second
order adiabatic particle number as 
\begin{equation}
{\cal N}_{k}^{(2)}\approx\frac{(A^{\prime})^{2}}{A^{4}}\frac{1}{k^{2}},\label{Number-adiabatic-1-UV}
\end{equation}
which depends on the isotropy-breaking factor $A(\tilde{\eta})$.
It is clear that, for modes with bigger mass, expression (\ref{Number-adiabatic-1-UV})
will contain further (positive) terms including $\tilde{m}\tilde{c}^{\prime}(\tilde{\eta})$.
So, number of created particles with bigger masses would be more abundant
than lighter ones. For massless modes, from Eq.~(\ref{Number-adiabatic-1})
we further obtain 
\begin{equation}
{\cal N}_{k}^{(0)}={\cal N}_{k}^{(2)}=0,\quad\quad\quad\quad{\cal N}_{k}^{(4)}=\frac{\big(\epsilon_{2}^{(2)}\big)^{2}}{16}=\frac{{\cal Q}^{2}}{16k^{4}},\label{Number-adiabatic-1-massless}
\end{equation}
with ${\cal Q}=-\tilde{a}^{\prime\prime}/\tilde{a}$. This indicates
that, in this case, no particle creation would occur up to the second
order in adiabatic series. However, in the fourth order, a small amount
of particles will be created. Therefore, in the adiabatic regime,
more massive particles would be created compared to the massless ones.
This expression is consistent with the discussion we had at the end
of section \ref{pairprod}.

The energy density of created particle for each mode is obtained from
Eqs. (\ref{energy-part-tot1}) and (\ref{Number-adiabatic-1}) as
\begin{align}
\varrho_{k}= &\ \tilde{\omega}_{k,\tilde{\eta}}\Big({\cal N}_{k}^{(0)}+{\cal N}_{k}^{(2)}+{\cal N}_{k}^{(4)}\Big).\nonumber \\
= &\ \frac{1}{16}\left[\frac{(\tilde{\omega}_{k,\tilde{\eta}}^{\prime})^{2}}{\tilde{\omega}_{k,\tilde{\eta}}^{3}}+\tilde{\omega}_{k,\tilde{\eta}}\big(\epsilon_{2}^{(2)}\big)^{2}-\frac{(\tilde{\omega}_{k,\tilde{\eta}}^{\prime})^{2}}{2\tilde{\omega}_{k,\tilde{\eta}}^{3}}\epsilon_{2}^{(2)}+\frac{\tilde{\omega}_{k,\tilde{\eta}}^{\prime}}{\tilde{\omega}_{k,\tilde{\eta}}^{2}}\epsilon_{2}^{\prime(3)}\right]\nonumber \\
= &\ \varrho_{k}^{(0)}+\varrho_{k}^{(2)}+\varrho_{k}^{(4)}.
\end{align}
By substituting this in Eq.~(\ref{energy-part-tot}), it is clear
that since $\varrho_{k}$ does not fall off faster than $k^{-4}$
when $k\rightarrow\infty$, the total energy density of (adiabatic)
particle productions would diverge. Since the zeroth adiabatic order
term in the energy density is zero, $\varrho_{k}^{(0)}=0$, the divergences
are included in the second and fourth order terms (for massive modes).
For massless modes, energy density of created particles is given only
by the fourth order term. So, the divergence is included only up to
fourth order terms. Therefore, the renormalized energy density of
created particles can be obtained by subtraction of the adiabatic
vacuum energy of the particle productions up to fourth order. Thus,
no back-reaction from created particles is present to generate any
inconsistency (i.e., producing an energy density $\rho_{{\rm par}}$
for scalar particles comparable to that of the background in Planck
regime) in studying of the quantum field and consequently analyzing
the particle productions on the background quantum geometry.

\section{Discussion and Outlook\label{conclusion}}

In this paper, following the strategy proposed in \citet{Ashtekar:2009mb},
we have studied the propagation of a scalar (inhomogeneous) quantum
test field $\varphi$, propagating over a quantized FLRW geometry
corresponding the to the Planck era. The geometry is coupled to a
massless homogeneous field $\phi$ which serves as the internal physical
time. In addition, we also compute the corresponding particle production
associated to this test field, when the spacetime transitions from
the quantum regime into the later classical regimes. 

More precisely, due to Planck era quantum gravitational effects, anisotropies
are induced in the the effective regime and thus the effective FLRW
geometry at the very early universe resembles a Biachi I model. Thus,
the quantum modes of the test field feel an effective geometry that
is anisotropic. We take the background geometry as the heavy degree
of freedom and each mode of the test field $\varphi$ as the light
(perturbation) degree of freedom. By neglecting backreactions, one
can consider a decomposition of the full state $\Psi(\nu,\mathsf{q}_{\mathbf{k}},\phi)$
of geometry plus matter, as a tensor product $\Psi(\nu,\mathsf{q}_{\mathbf{k}},\phi)=\Psi_{o}(\nu,\phi)\otimes\psi(\mathsf{q}_{\mathbf{k}},\phi)$,
where, the background wave function $\Psi_{o}$ (itself corresponding
to the tensor product of the geometry and the internal time $\phi$)
represents the probability amplitude for occurrence of the various
background FLRW geometries, and $\psi$ is the wave function of the
test field $\varphi$. One can then work in an interaction picture
to find an evolution equation (see Eq. \eqref{evol-eq1}) for $\psi(\mathsf{q}_{\mathbf{k}},\phi)$
with respect to the relational time $\phi$.

By matching and comparing the resulting effective evolution equation
for the state $\psi(\mathsf{q}_{\mathbf{k}},\phi)$ of massive (and
also massless) modes of $\varphi$ on the quantized background, and
the evolution equation of $\psi$ on a classical of Bianchi type I
spacetime, several classes of solutions were found for the test field
propagating on this ``dressed'' effective background geometry: 
\begin{enumerate}
\item For \emph{massless} modes of the test field, the dressed background
has an isotropic and homogeneous form with a dressed scale factor
$\tilde{a}(\phi)$ given by \eqref{massless-dressed-scale-1}, which
is a function of the expectation values of operators of the corresponding
quantum geometry (see similar solution in Ref.~\cite{Ashtekar:2009mb}).
\item For \emph{massive} modes, we have considered two cases: dressed and
undressed mass. In both cases the emergent dressed background is the
Bianchi type-I geometry with the scale factors $\tilde{a}_{1},\tilde{a}_{2},\tilde{a}_{3}$.
\end{enumerate}
Next, in section \ref{pairprod}, we have investigated the issue of
particle production associated to the test field, as a result of transitioning
from the effective Bianchi I geometry to a classical isotropic spacetime.
This can provide a means to measure some of the interesting QFT-related
phenomena from quantum gravity regime in Planck era (see Ref. \cite{Tavakoli:2014mra}
for the issue of particle creation phenomena in a bouncing scenario
of LQC). Parker has shown that while the particle production does
not happen for massless conformally coupled fields, it is generic
for massive scalar fields of arbitrary coupling to the spacetime curvature.
Typically, the particle production is significant when the particle
mass $m$ is of the order of the expansion rate $H$, or when $m^{2}\sim\dot{H}$.
Simple dimensional arguments show that in a Universe with critical
density $\rho_{c}\sim3H^{2}/8\pi G$, the quantum particle production
can contribute significantly to the energy density of the Universe
at early epochs when $H$ is not too far below the Planck scale. We
have shown that, in our model, as long as the quantum conditions \eqref{condition-1}
and \ref{condition-2} are valid, particle creation always occurs
in transition from quantum to classical regime.

In section \ref{fluctuation}, we generalized our quantum theory of
single mode to the theory of a quantum field, including all possible
modes, which propagate on the dressed anisotropic geometry we have
derived in the previous section. More precisely, for a given solution of the anisotropic dressed background, we assumed that all
modes feel the same  geometry, thereby, we studied the quantum theory of all modes and looked for possible observable signatures stemming from the early stages of the Universe.
In particular,
we have investigated the amplitude of quantum fluctuations on the
emergent effective anisotropic background which can be a candidate
to study the origin of the anisotropy seen from observation of CMB.
We provided a relation for the power spectrum of the field on the
anisotropic, dressed geometry and compared our result to that of the
isotropic case. We have shown that the power spectrum of the anisotropic
case is different from the isotropic one and have provided an expression
that reflects this difference. This would imprint a signature on the
power spectrum of the CMB \cite{Komatsu:2010fb}, distinct from the
one associated to the isotropic dressed, background.

It is worth noting that there are other models such as the Belinskii,
Khalatnikov and Lifshitz (BKL) cosmology \cite{Belinsky:1970ew,Belinsky:1982pk}
that bare some resemblance to the LQC while there are also important
differences between them. First off, BKL is mostly a classical model
while LQC is based on quantum gravity. In both models the Universe
is oscillating. However, there exists a big bang singularity in the
BKL model which disappears in LQC due to quantum effects involved.
In fact, in LQC the Friedmann equation effectively acquires an additional
``repulsive'' term that only dominates near the Planck regime, where
it ``pushes'' the Universe ``outwards'', overcoming the classical
gravitational attraction. Also while in the BKL model the spatial
points decouple from each other near singularity and the Universe
behaves chaotically, LQC drastically changes this chaotic behavior
which is again due to the quantum effects it incorporates into the
analysis. Nevertheless, the analysis of the fate of generic space-like
singularities due to the BKL conjecture, in the complete picture of
LQG is still under investigation \cite{Ashtekar:2008jb,Ashtekar:2011ck}.

From previous studies we know that there are other curious effects
associated to the propagation of massless scalar fields and vector
fields, on an effective spacetime in four or lower dimensions \cite{Gambini:2009ie,Gambini:2011mw,Gambini:2011nx,Bonder:2017ckx,Corichi:2015vsa}.
As a future work, it would be interesting to study similar effects
corresponding to the massive scalar field and the connection of these
effects to cosmological phenomena and observations. 
\begin{acknowledgments}
The work of YT was supported by the Polish Narodowe Centrum Nauki
(NCN) grant 2012/05/E/ST2/03308. He also thanks Brazilian agency FAPES/CAPES
and INEF (Iran) for partial financial supports. JCF thanks Brazilian
agencies CNPq and FAPES for partial financial support. This article
is based upon work from European Cooperation in Science and Technology
(COST) action CA18108 --- Quantum gravity phenomenology in the multi-messenger
approach, supported by COST. 
\end{acknowledgments}

\appendix

\section{Fourier expansion of the inhomogeneous field\label{field-equation-detail}}

Let us write the field $\varphi(x_{0},\mathbf{x})$ as a Fourier decomposition
of the form 
\begin{equation}
\varphi(x_{0},\mathbf{x})=\frac{1}{(2\pi)^{3/2}}\sum_{\mathbf{k}\in{\cal L}}\varphi_{\mathbf{k}}(x_{0})e^{i\mathbf{k}\cdot\mathbf{x}},\label{total-field1a}
\end{equation}
where ${\cal L}$ was defined in Eq. \eqref{lattice}. By decomposition
of the field modes into the real and imaginary parts we have 
\begin{equation}
\varphi_{\mathbf{k}}(x_{0})=\frac{1}{\sqrt{2}}\left[\varphi_{\mathbf{k}}^{(1)}(x_{0})+i\varphi_{\mathbf{k}}^{(2)}(x_{0})\right].\label{decomposition-field}
\end{equation}
Reality condition, $\varphi^{\ast}(x_{0},\mathbf{x})=\varphi(x_{0},\mathbf{x})$,
implies that 
\begin{equation}
\varphi_{\mathbf{k}}^{(1)}=\varphi_{-\mathbf{k}}^{(1)}\quad{\rm and}\quad\varphi_{-\mathbf{k}}^{(2)}=-\varphi_{\mathbf{k}}^{(2)}.
\end{equation}
By introducing new variables $\mathsf{q}_{\pm\mathbf{k}}$ associated
to the real variables $\varphi_{\mathbf{k}}^{(1)}$ and $\varphi_{\mathbf{k}}^{(2)}$
for the positive and negative sections of ${\cal L}$ as \cite{Ashtekar:2009mb}
\begin{equation}
\mathsf{q}_{\mathbf{k}}(x_{0})=\begin{cases}
\varphi_{\mathbf{k}}^{(1)}(x_{0}), & \text{if \ensuremath{\mathbf{k}\in{\cal L}_{+}}},\\
\\
\varphi_{-\mathbf{k}}^{(2)}(x_{0}), & \text{if \ensuremath{\mathbf{k}\in{\cal L}_{-}}},
\end{cases}\label{q-variable}
\end{equation}
we can rewrite the field modes $\varphi_{\mathbf{k}}$ for each $\mathbf{k}\in{\cal L}_{+}$,
in terms of $\mathsf{q}_{\pm\mathbf{k}}$ as 
\begin{equation}
\varphi_{\mathbf{k}}(x_{0})=\frac{1}{\sqrt{2}}\left[\mathsf{q}_{\mathbf{k}}(x_{0})+i\mathsf{q}_{-\mathbf{k}}(x_{0})\right].\label{decomposition-field2}
\end{equation}
The momentum $\pi_{\varphi}(x_{0},\mathbf{x})$ conjugate to $\varphi(x_{0},\mathbf{x})$
can be Fourier decomposed as 
\begin{equation}
\pi_{\varphi}(x_{0},\mathbf{x})=\frac{1}{(2\pi)^{3/2}}\sum_{\mathbf{k}\in{\cal L}}\pi_{\mathbf{k}}(x_{0})e^{i\mathbf{k}\cdot\mathbf{x}}.\label{total-field1-a}
\end{equation}
Then separating the momentum $\pi_{\mathbf{k}}$ associated to each
mode $\mathbf{k}\in{\cal L}$, into the real and imaginary parts gives
\begin{equation}
\pi_{\mathbf{k}}(x_{0})=\frac{1}{\sqrt{2}}\left[\pi_{\mathbf{k}}^{(1)}(x_{0})+i\pi_{\mathbf{k}}^{(2)}(x_{0})\right].\label{decomposition-momentum}
\end{equation}
In terms of the new real variables $\mathsf{p}_{\pm\mathbf{k}}$ (conjugate
to $\mathsf{q}_{\pm\mathbf{k}}$), defined by \cite{Ashtekar:2009mb,Dapor:2012jg}
\begin{equation}
\mathsf{p}_{\mathbf{k}}(x_{0})=\begin{cases}
\pi_{\mathbf{k}}^{(1)}(x_{0}), & \text{if \ensuremath{\mathbf{k}\in{\cal L}_{+}}},\\
\\
\pi_{-\mathbf{k}}^{(2)}(x_{0}), & \text{if \ensuremath{\mathbf{k}\in{\cal L}_{-}}}\,,
\end{cases}\label{p-variable}
\end{equation}
we rewrite the mode functions (\ref{decomposition-momentum}), for
each mode $\mathbf{k}\in{\cal L}_{+}$, as 
\begin{equation}
\pi_{\mathbf{k}}(x_{0})=\frac{1}{\sqrt{2}}\left[\mathsf{p}_{\mathbf{k}}(x_{0})+i\mathsf{p}_{-\mathbf{k}}(x_{0})\right].\label{decomposition-momentum2}
\end{equation}
The conjugate variables $(\mathsf{q}_{\mathbf{k}},\mathsf{p}_{\mathbf{k}})$
defined in Eqs.~(\ref{q-variable}) and (\ref{p-variable}) are in
fact the variables we have used for the field in formulating the Hamiltonian
in Eq. (\ref{Hamiltonian-SF1}), but of course by assuming a discrete
lattice ${\cal L}$ (see Ref. \cite{Ashtekar:2009mb}) rather than
continuous space for the modes we have considered in studying of field
theory in section \ref{fluctuation}.

By defining an auxiliary field $\chi_{\mathbf{k}}=\tilde{c}^{1/2}\varphi_{\mathbf{k}}$,
and decomposing it into the real and imaginary parts as in (\ref{decomposition-field}),
in terms of the variable $\mathsf{q}_{\pm\mathbf{k}}(x_{0})$, we
have 
\begin{equation}
\chi_{\mathbf{k}}(x_{0})=\frac{\tilde{c}^{1/2}}{\sqrt{2}}\left[\mathsf{q}_{\mathbf{k}}(x_{0})+i\mathsf{q}_{-\mathbf{k}}(x_{0})\right],\label{decomposition-fieldax}
\end{equation}
for  each $\mathbf{k}\in{\cal L}_{+}$. By substituting $\mathsf{q}_{\mathbf{k}}$
from Eq.~(\ref{decomposition-fieldax}) into  Eq.~(\ref{motion1-mod})
in terms of $\chi_{\mathbf{k}}$, and rewriting now the derivation
with respect to the conformal time $\tilde{\eta}$, defined in Eq.~(\ref{conformaltime}),
we obtain the field equation (\ref{field-eq-mode1-2a}) for the dressed
Bianchi type I geometry (\ref{BI-metric-class2}) emerged for our
background spacetime.

By comparing Eq.~(\ref{field-eq-mode1-2a}) with the field equation
on a Bianchi type I spacetime in Ref.~\cite{Birrell:1982ix}: 
\begin{equation}
\chi_{\mathbf{k}}^{\prime\prime}+\Big[\tilde{c}(\tilde{\eta})\Big(\sum_{i}\frac{\tilde{k}_{i}^{2}}{\tilde{c}_{i}}+\tilde{m}^{2}-\frac{\tilde{{\cal R}}}{6}\Big)+Q\Big]\chi_{\mathbf{k}}=0,
\end{equation}
the Ricci scalar $\tilde{{\cal R}}$ for the herein dressed Bianchi
geometry becomes 
\begin{equation}
\tilde{{\cal R}}=3\tilde{c}^{\prime\prime}\tilde{c}^{-2}-\frac{3}{2}\tilde{c}^{\prime2}\tilde{c}^{-3}-6\tilde{c}^{-1}Q,
\end{equation}
where we have defined 
\begin{equation}
d_{i}\coloneqq\frac{\tilde{c}_{i}^{\prime}}{\tilde{c}_{i}}\quad\quad{\rm and}\quad\quad Q\coloneqq\frac{1}{72}\sum_{i<j}(d_{i}-d_{j})^{2}.\label{c-tilde-2a}
\end{equation}

\section{Fourth order adiabatic expansion\label{Ad-Vac}}

We notice that ${\cal Q}(\tilde{\eta})=\tilde{c}^{\prime\prime}/2\tilde{c}-\tilde{c}^{2}/4\tilde{c}^{2}$
is second order in time derivative of the metric components, while
$X\equiv\tilde{\omega}_{k,\tilde{\eta}}^{2}$ is of zeroth order.
From (\ref{omega-BI-1}) we have that $Y=X-{\cal Q}$, so $Y$ contains
zeroth and second order terms. Next, from Eqs. (\ref{epsilon-under-1}),
(\ref{epsilon-under-2}), (\ref{adibatic-2-sec1}) and (\ref{adibatic-2-sec})
we can rewrite $\epsilon_{2}$ and $\epsilon_{4}$ as 
\begin{align}
\epsilon_{2}= & -\frac{1}{4}Y^{-2}Y^{\prime\prime}+\frac{5}{16}Y^{-3}(Y^{\prime})^{2}-X^{-1}{\cal Q},\label{adibatic-2-sec1App}\\
\epsilon_{4}= & -\frac{1}{4}Y^{-1}(1+\underline{\epsilon_{2}})^{-2}\Big[\underline{\epsilon_{2}}^{\prime\prime}-\frac{1}{2}Y^{-1}Y^{\prime}\underline{\epsilon_{2}}^{\prime}-\frac{5}{4}(1+\underline{\epsilon_{2}})^{-1}(\underline{\epsilon_{2}}^{\prime})^{2}\Big]-\underline{\epsilon_{2}}X^{-1}{\cal Q}.\label{adibatic-2-sec-App}
\end{align}
where 
\begin{equation}
\underline{\epsilon_{2}}=-\frac{1}{4}Y^{-2}Y^{\prime\prime}+\frac{5}{16}Y^{-3}(Y^{\prime})^{2}.%
\label{eps-1}
\end{equation}
It is clear that $\underline{\epsilon_{2}}$ contains terms of second,
fourth and higher orders in time derivative of the metric components.
Then, the first derivative of $\underline{\epsilon_{2}}$ contains
third and higher order terms. Consequently, $\underline{\epsilon_{2}}^{\prime\prime}$
contains terms of orders equal to and bigger than four. Therefore,
$\epsilon_{2}$ contains second and higher order terms. Using these
in Eq.~(\ref{adibatic-2-sec-App}) we conclude that the leading order
term in $\epsilon_{4}$ is of order four. Now, we decompose $\epsilon_{2}$
(up to the fourth order), as 
\begin{equation}
\epsilon_{2}=\epsilon_{2}^{(2)}+\epsilon_{2}^{(4)}+{\rm Higher~order~terms},\label{adibatic-2-sec1App-1}
\end{equation}
where 
\begin{align}
\epsilon_{2}^{(2)}= &\ -\frac{1}{4}X^{-2}X^{\prime\prime}+\frac{5}{16}X^{-3}(X^{\prime})^{2}-X^{-1}{\cal Q},\label{adibatic-2-sec1App-1a}\\
\epsilon_{2}^{(4)}= &\ \frac{1}{4}X^{-2}{\cal Q}^{\prime\prime}-\frac{1}{2}X^{-3}X^{\prime\prime}{\cal Q}-\frac{5}{8}X^{-3}X^{\prime}{\cal Q}^{\prime}+\frac{15}{16}X^{-4}(X^{\prime})^{2}{\cal Q}.\label{adibatic-2-sec1App-1b}
\end{align}
Similarly, decomposition of $\epsilon_{4}$ in terms of fourth and
higher order terms reads 
\begin{align}
\epsilon_{4}= &\ \epsilon_{4}^{(4)}~+~{\rm Higher~order~terms}\nonumber \\
= &\ -\frac{1}{4}X^{-1}\Big[\underline{\epsilon_{2}}^{\prime\prime}-\frac{1}{2}\underline{\epsilon_{2}}^{\prime}X^{-1}X^{\prime}\Big]-\underline{\epsilon_{2}}X^{-1}{\cal Q}\nonumber \\
= &\ \frac{1}{16}X^{\prime\prime\prime\prime}X^{-3}-\frac{1}{8}X^{\prime\prime\prime}X^{\prime}X^{-4}+\frac{3}{8}(X^{\prime})^{2}X^{\prime\prime}X^{-5}+\frac{5}{64}\Big[3X^{\prime\prime}-4(X^{\prime})^{2}X^{-1}\Big](X^{\prime})^{2}X^{-5}\nonumber \\
 & -\frac{1}{8}X^{-1}\Big((X^{\prime\prime})^{2}+X^{\prime}X^{\prime\prime\prime}\Big)\Big(X^{-3}+\frac{5}{4}\Big)-\frac{1}{32}X^{\prime\prime\prime}X^{-4}X^{\prime}+\frac{1}{16}X^{-2}(X^{\prime})^{2}X^{\prime\prime}\Big(X^{-3}+\frac{5}{4}\Big)\nonumber \\
 & +\frac{1}{4}\Big[X^{\prime\prime}-\frac{5}{4}X^{-1}(X^{\prime})^{2}\Big]X^{-3}{\cal Q}-\frac{5}{128}(X^{\prime})^{4}X^{-6}+~{\rm Higher~order~terms}.\label{adibatic-2-sec-App-2}
\end{align}

\section{Renormalization of the energy-momentum tensor\label{adiabatic-energy-momentum}}

A key step in constructing a theory of quantum fields is identifying
the energy-momentum tensor of the quantized field, which is presumably
obtainable from the divergent expression for the energy-momentum tensor
that results from the formal Lagrangian field theory. In this appendix,
we follow the method of Ref.~\cite{Parker:1974qw} by applying the
``adiabatic regularization'' to the anisotropic dressed metric (\ref{BI-metric-class2}).
The divergences in the energy-momentum tensor are isolated in the
three leading terms of asymptotic expansion corresponding to an adiabatic
limit, i.e., the limit of slow time dependence of the metric (\ref{BI-metric-class2}).
The quantity expanded is the expectation value of the energy-momentum
tensor with respect to the approximate vacuum state associated to
the mode function defined in (\ref{adibatic-1}).

The energy-momentum tensor of a minimally coupled, massive field (\ref{total-field1})
is given by 
\begin{equation}
T_{ab}=\nabla_{a}\varphi\nabla_{b}\varphi-\frac{1}{2}\tilde{g}_{ab}\big(\tilde{g}^{cd}\nabla_{c}\varphi\nabla_{d}\varphi+m^{2}\varphi^{2}\big).\label{Energy-Mom-class}
\end{equation}
Let $|\tilde{0}\rangle$ be a normalized vacuum state annihilated
by all $\hat{a}_{\mathbf{k}}$. With respect to this vacuum state,
the formal expression for the expectation value of the energy density
operator, $\langle\tilde{0}|\hat{\rho}|\tilde{0}\rangle=-\langle\tilde{0}|\hat{T}_{0}^{0}|\tilde{0}\rangle$,
of the field $\varphi(\tilde{\eta},\mathbf{x})$ on the dressed spacetime
(\ref{BI-metric-class2}) is obtained as 
\begin{equation}
\langle\tilde{0}|\hat{\rho}|\tilde{0}\rangle \eqqcolon \frac{\hbar}{\ell^{3}\tilde{c}^{2}}\sum_{\mathbf{k}}\rho_{k}[u_{k}(\tilde{\eta})],\label{energy-tot-vac1}
\end{equation}
where $\rho_{k}$ is given by 
\begin{equation}
\rho_{k}=\frac{1}{4}\Big[|u_{k}^{\prime}|^{2}+\Big(\tilde{\omega}_{\tilde{\eta},k}^{2}+\frac{1}{4}\frac{\tilde{c}^{\prime2}}{\tilde{c}^{2}}\Big)|u_{k}|^{2}-\frac{\tilde{c}^{\prime}}{2\tilde{c}}\big(u_{k}^{\ast}u_{k}^{\prime}+u_{k}u_{k}^{\prime\ast}\big)\Big].\label{energy-summand}
\end{equation}
In this formalism, the quantum theory of test field, reduces to a
consideration of the classical wave equation (\ref{KG-1-eff-2}) for
$u_{k}$. The expression $\langle\tilde{0}|\hat{\rho}|\tilde{0}\rangle$
is infinite and must be renormalized by subtracting the divergent
terms. Therefore, in the following, we introduce the adiabatic regularization
technique in order to handle with divergences of the energy and pressure
components of the quantized field.

For each $\underline{u}_{k}$ in the summand (\ref{energy-tot-vac1}),
it can be shown that all ultra-violet divergences are contained in
terms of adiabatic order equal to and smaller than four. Our task
now is to expand the energy and pressure for each mode $k$ in asymptotic
series, by using adiabatic approximation given in (\ref{adibatic-2}).

We start from the (adiabatic) regularization of energy density operator
(\ref{energy-summand}). We compute $\langle\tilde{0}|\hat{\rho}|\tilde{0}\rangle_{{\rm ren}}$,
up to the fourth adiabatic order, as 
\begin{equation}
\langle\tilde{0}|\hat{\rho}|\tilde{0}\rangle_{{\rm ren}}=\frac{\hbar}{\ell^{3}\tilde{c}^{2}}\sum_{\mathbf{k}}\Big(\rho_{k}[u_{k}(\tilde{\eta})]-\underline{\rho_{k}}(\tilde{\eta})\Big),
\end{equation}
where the subtraction term $\underline{\rho_{k}}(\tilde{\eta})\equiv\rho_{k}[\underline{u}_{k}(\tilde{\eta})]$
(being the adiabatic vacuum energy for each mode) is needed to regularize
the energy density. Here $\underline{\rho_{k}}$ is obtained by the
terms of zeroth, second and fourth adiabatic orders in the expansion
of the summand as 
\begin{equation}
\underline{\rho_{k}}(\tilde{\eta})=\rho_{k}^{(0)}+\rho_{k}^{(2)}+\rho_{k}^{(4)}.\label{C-rho}
\end{equation}
To the fourth order we have 
\begin{align}
|\underline{u}_{k}|^{2}= &\ \big(W_{k}(\tilde{\eta})\big)^{-1}=\tilde{\omega}_{\tilde{\eta},k}^{-1}\Big[1-\frac{1}{2}\big(\epsilon_{2}^{(2)}+\epsilon_{2}^{(4)}+\epsilon_{4}^{(4)}\big)+\frac{3}{8}\big(\epsilon_{2}^{(2)}\big)^{2}\Big],\label{v-vprime1}\\
|\underline{u}_{k}^{\prime}|^{2}= &\ W_{k}(\tilde{\eta})+\frac{1}{4}\big(W_{k}(\tilde{\eta})\big)^{-3}\big(W_{k}^{\prime}(\tilde{\eta})\big)^{2}\nonumber \\
= &\ \tilde{\omega}_{\tilde{\eta},k}\Big[1+\frac{1}{2}\big(\epsilon_{2}^{(2)}+\epsilon_{2}^{(4)}+\epsilon_{4}^{(4)}\big)-\frac{1}{8}\big(\epsilon_{2}^{(2)}\big)^{2}\Big]\nonumber \\
 & \quad\quad\quad+\frac{1}{4}\tilde{\omega}_{\tilde{\eta},k}^{-3}(\tilde{\omega}_{\tilde{\eta},k}^{\prime})^{2}\Big[1-\frac{1}{2}\epsilon_{2}^{(2)}\Big]+\frac{1}{4}\tilde{\omega}_{\tilde{\eta},k}^{\prime}\tilde{\omega}_{\tilde{\eta},k}^{-2}\epsilon_{2}^{\prime(3)},\label{v-vprime2}
\end{align}
and 
\begin{equation}
2\Re(\underline{u}_{k}\underline{u}_{k}^{\prime\ast})=-W_{k}^{\prime}(\tilde{\eta})\big(W_{k}(\tilde{\eta})\big)^{-2}=-\tilde{\omega}_{\tilde{\eta},k}^{-1}\Big[\frac{\tilde{\omega}_{\tilde{\eta},k}^{\prime}}{\tilde{\omega}_{\tilde{\eta},k}}\Big(1-\frac{1}{2}\epsilon_{2}^{(2)}\Big)+\frac{1}{2}\epsilon_{2}^{\prime(3)}\Big],\label{v-vprime3}
\end{equation}
where, $2\Re(\underline{u}_{k}\underline{u}_{k}^{\prime\ast})=\big(\underline{u}_{k}^{\ast}\underline{u}_{k}^{\prime}+\underline{u}_{k}\underline{u}_{k}^{\prime\ast}\big)$.
By replacing Eqs.~(\ref{v-vprime1}) and (\ref{v-vprime2}) and (\ref{v-vprime3})
in (\ref{energy-summand}), we obtain the zeroth, second and fourth
adiabatic order of vacuum energy density as 
\begin{eqnarray}
\rho_{k}^{(0)} &= & \frac{\tilde{\omega}_{\tilde{\eta},k}}{2},\label{energy-ren-0}\\
\rho_{k}^{(2)} &= & \frac{1}{16}\frac{(\tilde{\omega}_{\tilde{\eta},k}^{\prime})^{2}}{\tilde{\omega}_{\tilde{\eta},k}^{3}}+\frac{1}{16\tilde{\omega}_{\tilde{\eta},k}}\frac{\tilde{c}^{\prime2}}{\tilde{c}^{2}}+\frac{\tilde{c}^{\prime}}{8\tilde{c}}\frac{\tilde{\omega}_{\tilde{\eta},k}^{\prime}}{\tilde{\omega}_{\tilde{\eta},k}^{2}}\,,\label{energy-ren-2}\\
\rho_{k}^{(4)} &= & \frac{\tilde{\omega}_{\tilde{\eta},k}}{16}\big(\epsilon_{2}^{(2)}\big)^{2}+\frac{1}{16}\Big(\frac{\tilde{\omega}_{\tilde{\eta},k}^{\prime}}{\tilde{\omega}_{\tilde{\eta},k}^{2}}+\frac{1}{\tilde{\omega}_{\tilde{\eta},k}}\frac{\tilde{c}^{\prime}}{\tilde{c}}\Big)\epsilon_{2}^{\prime(3)}-\frac{1}{32}\Big(\frac{(\tilde{\omega}_{\tilde{\eta},k}^{\prime})^{2}}{\tilde{\omega}_{\tilde{\eta},k}^{3}}+\frac{1}{\tilde{\omega}_{\tilde{\eta},k}}\frac{\tilde{c}^{\prime2}}{\tilde{c}^{2}}+2\frac{\tilde{c}^{\prime}}{\tilde{c}}\frac{\tilde{\omega}_{\tilde{\eta},k}^{\prime}}{\tilde{\omega}_{\tilde{\eta},k}^{2}}\Big)\epsilon_{2}^{(2)}.\quad\quad\quad\label{energy-ren-4}
\end{eqnarray}
The expressions for $\epsilon_{2}^{(2)}$ and $\epsilon_{2}^{(4)}$
are given by Eqs.~(\ref{adibatic-2-sec1App-1a}) and (\ref{adibatic-2-sec1App-1b}).
Notice that, the terms (\ref{energy-ren-0})-(\ref{energy-ren-4})
are state-independent and local in the background geometry. The significance
of Eq.~(\ref{C-rho}) is that the ultra-violet divergences associated
with the zeroth, second and fourth-order terms can be removed by renormalization.

The formal expression for the expectation value of the pressure operator
is given by 
\begin{equation}
\langle\tilde{0}|\hat{P}_{i}|\tilde{0}\rangle\coloneqq\frac{\hbar}{\ell^{3}\tilde{c}^{2}}\sum_{\mathbf{k}}P_{i,k}[u_{k}(\tilde{\eta})],\label{stress-class-3B}
\end{equation}
where, for each mode $u_{k}$ we have 
\begin{equation}
P_{i,k}=\frac{\tilde{c}_{i}}{4}\Big[|u_{k}^{\prime}|^{2}-\Big(\tilde{\omega}_{\tilde{\eta},k}^{2}-\frac{1}{4}\frac{\tilde{c}^{\prime2}}{\tilde{c}^{2}}-2k_{i}^{2}\frac{\tilde{c}}{\tilde{c}_{i}}\Big)|u_{k}|^{2}-\frac{1}{2}\frac{\tilde{c}^{\prime}}{\tilde{c}}\big(u_{k}u_{k}^{\ast\prime}+u_{k}^{\prime}u_{k}^{\ast}\big)\Big].\label{stress-class-3B-2}
\end{equation}
The renormalized vacuum expectation value of the stress operator is
obtained by subtracting the divergent terms as 
\begin{equation}
\langle\tilde{0}|\hat{P}_{i}|\tilde{0}\rangle_{{\rm ren}}=\frac{\hbar}{\ell^{3}\tilde{c}^{2}}\sum_{\mathbf{k}}\Big(P_{i,k}[u_{k}(\tilde{\eta})]-\underline{P_{i,k}}(\tilde{\eta})\Big).\label{P-ren1}
\end{equation}
Here, the subtraction term $\underline{P_{i,k}}(\tilde{\eta})\equiv P_{i,k}[\underline{u}_{k}(\tilde{\eta})]$
associated to the vacuum expectation value, is obtained, up to the
fourth adiabatic order, as 
\begin{equation}
\underline{P_{i,k}}(\tilde{\eta})=P_{i,k}^{(0)}+P_{i,k}^{(2)}+P_{i,k}^{(4)}.\label{P-ren2}
\end{equation}
By a similar analysis as for the energy density operator, the zeroth,
second and fourth order terms of $\underline{P_{i,k}}$ on the right
hand side of Eq.~(\ref{P-ren2}) are given by 
\begin{align}
P_{i,k}^{(0)}= &\ \frac{\tilde{c}}{2\tilde{\omega}_{\tilde{\eta},k}}\tilde{k}_{i}^{2},\label{P-ren1app}\\
P_{i,k}^{(2)}= &\ \frac{\tilde{c}_{i}}{4\tilde{\omega}_{\tilde{\eta},k}}\Big[\frac{\tilde{c}^{\prime2}}{4\tilde{c}^{2}}+\frac{(\tilde{\omega}_{\tilde{\eta},k}^{\prime})^{2}}{4\tilde{\omega}_{\tilde{\eta},k}^{2}}+\frac{\tilde{c}^{\prime}}{2\tilde{c}}\frac{\tilde{\omega}_{\tilde{\eta},k}^{\prime}}{\tilde{\omega}_{\tilde{\eta},k}}+\Big(\tilde{\omega}_{\tilde{\eta},k}^{2}-\tilde{k}_{i}^{2}\frac{\tilde{c}}{\tilde{c}_{i}}\Big)\epsilon_{2}^{(2)}\Big],\label{P-ren2app}\\
P_{i,k}^{(4)}= &\ \frac{\tilde{c}_{i}}{4\tilde{\omega}_{\tilde{\eta},k}}\Big[\Big(\tilde{\omega}_{\tilde{\eta},k}^{2}-\tilde{k}_{i}^{2}\frac{\tilde{c}}{\tilde{c}_{i}}\Big)\big(\epsilon_{2}^{(4)}+\epsilon_{4}^{(4)}\big)+\Big(\frac{3}{4}\tilde{k}_{i}^{2}\frac{\tilde{c}}{\tilde{c}_{i}}-\frac{\tilde{\omega}_{\tilde{\eta},k}^{2}}{2}\Big)\big(\epsilon_{2}^{(2)}\big)^{2}\nonumber \\
 & \quad\quad\quad\quad\quad+\Big(\frac{\tilde{\omega}_{\tilde{\eta},k}^{\prime}}{4\tilde{\omega}_{\tilde{\eta},k}}+\frac{\tilde{c}^{\prime}}{4\tilde{c}}\Big)\epsilon_{2}^{\prime(3)}-\Big(\frac{(\tilde{\omega}_{\tilde{\eta},k}^{\prime})^{2}}{8\tilde{\omega}_{\tilde{\eta},k}^{2}}+\frac{\tilde{c}^{\prime2}}{8\tilde{c}^{2}}+\frac{\tilde{c}^{\prime}}{4\tilde{c}}\frac{\tilde{\omega}_{\tilde{\eta},k}^{\prime}}{\tilde{\omega}_{\tilde{\eta},k}}\Big)\epsilon_{2}^{(2)}\Big].\label{P-ren3app}
\end{align}
The formal expression for the vacuum expectation value of the Hamiltonian
operator $\hat{H}_{\varphi}$, for the quantized massive field, on
the effective background (\ref{BI-metric-class2}) is obtained as
\begin{equation}
\langle\tilde{0}|\hat{H}_{\varphi}|\tilde{0}\rangle=\frac{\hbar}{\ell^{3}}\sum_{\mathbf{k}}\tilde{H}_{k}[u_{k}(\tilde{\eta})],\label{Hamilton-conformal-1}
\end{equation}
where, $\tilde{H}_{k}$ is the expectation value of the Hamiltonian
for $k$th mode. By setting $\tilde{N}_{x_{0}}=\tilde{c}^{1/2}(\tilde{\eta})$
we achieve the expression for $\tilde{H}_{k}$ in conformal time as
\begin{equation}
\tilde{H}_{k}(\tilde{\eta})=\frac{1}{4}\Big[|u_{k}^{\prime}|^{2}+\Big(\frac{1}{4}\frac{\tilde{c}^{\prime2}}{\tilde{c}^{2}}+\tilde{\omega}_{\tilde{\eta},k}^{2}\Big)|u_{k}|^{2}-\frac{1}{2}\frac{\tilde{c}^{\prime}}{\tilde{c}}\big(u_{k}u_{k}^{\ast\prime}+u_{k}^{\prime}u_{k}^{\ast}\big)\Big]=\rho_{k}(\tilde{\eta}).\label{Hamilton-conformal-2}
\end{equation}
Therefore, it is straightforward to show that the renormalized vacuum
expectation value of the field's Hamiltonian is 
\begin{align}
\langle\tilde{0}|\hat{H}_{\varphi}|\tilde{0}\rangle_{{\rm ren}}= &\ \frac{\hbar}{\ell^{3}}\sum_{\mathbf{k}}\Big(\tilde{H}_{k}[u_{k}(\tilde{\eta})]-\underline{H_{k}}(\tilde{\eta})\Big)\nonumber \\
= & \ \tilde{c}^{2}(\tilde{\eta})\langle\tilde{0}|\hat{\rho}|\tilde{0}\rangle_{{\rm ren}},\label{Hamiltonian-ren}
\end{align}
where $\underline{H_{k}}=\tilde{H}_{k}(\underline{u}_{k})$, is given,
up to the fourth order, by 
\begin{equation}
\underline{H_{k}}(\tilde{\eta})=\tilde{c}^{2}\big(\rho_{k}^{(0)}+\rho_{k}^{(2)}+\rho_{k}^{(4)}),
\end{equation}
The terms $\rho_{k}^{(n)}$ (with $n=1,2,3$) are given by Eqs.~(\ref{energy-ren-0}),
(\ref{energy-ren-2}) and (\ref{energy-ren-4}). Therefore, expectation
values of renormalized Hamiltonian operator (\ref{Hamiltonian-ren})
is well-defined on any fourth order adiabatic state.

\bibliography{Bibliography}

\end{document}